\documentclass[amssymb,amsmath,preprint,superscriptaddress,nofootinbib,pra,aps]{revtex4-1}

\usepackage{amsmath}
\usepackage{graphicx}
\usepackage{gensymb}
\usepackage{xcolor}
\usepackage{slashed}
\usepackage{graphicx}
\usepackage{amsmath}
\usepackage{amssymb}
\usepackage{tabularx}
\usepackage{array}
\usepackage{textcomp}
\usepackage[colorlinks]{hyperref}

\begin{document}
\title{Neutrino nucleus Quasi-Elastic and resonant Neutral Current scatterings with Non-Standard Interactions}

\author{S. Abbaslu}
\email{s-abbaslu@ipm.ir}
\affiliation{School of Physics, Institute for Research in Fundamental Sciences (IPM), Tehran, Iran}

\author{M. Dehpour}
\email{m.dehpour@cern.ch}
\affiliation{Faculty of Mathematics and Physics, Charles University, Prague, Czechia}

\author{Y. Farzan}
\email{yasaman@theory.ipm.ac.ir}
\affiliation{School of Physics, Institute for Research in Fundamental Sciences (IPM), Tehran, Iran}

\author{S. Safari}
\email{s.safari@uni-mainz.de}
\affiliation{Institute of Physics, Johannes Gutenberg University, Mainz, Germany}

\begin{abstract}
    As well known, the cross sections of the resonance and Quasi-Elastic (QE) scattering off nucleons depend on quantities known as form factors that describe the nucleon structure. There are alternative approaches to determine the values of these non-perturbative quantities, some of them relying on the Neutral Current (NC) scattering of neutrinos off nucleons. In the presence of NC Non-Standard Interactions (NSI), such derivations must be revisited.  The aim of the present paper is to discuss how information on NSI can be extracted by combining alternative approaches for deriving the form factors. We discuss how the KamLAND atmospheric neutrino data with $E_\nu<{\rm GeV}$ (used to determine the axial strange form factor $g_A^s$) can already constrain the axial NSI of $\nu_\tau$ with nucleons. We also argue that if the precision measurement of $\nu_\mu$ NC QE scattering establishes unexpectedly large vector strange form factor ({\it e.g.,} $F_1^s(Q^2)\sim 0.01$), it will be an indication for nonzero NSI coupling with $u$ and $d$ quarks  ($\epsilon_{\mu\mu}^{Au/d}\sim 0.01$). We study the QE and resonance scattering cross sections of $\nu_\tau$ and $\nu_e$ off Argon and show that if their axial NSI is of the order of (but of course below) the present bounds, the deviation of QE cross sections from the SM prediction will be sizable and distinct from the uncertainties induced by the form factors.
\end{abstract}

\maketitle

\section{Introduction}
\label{sec:intro}
Studying the interaction of neutrinos with matter fields has been pivotal in building the Standard Model (SM) of particle physics. Indeed the establishment of the ${\rm SU}(2)\times {\rm U}(1)$ gauge symmetry as the basis of electroweak interaction was made possible by the discovery of NC interaction between neutrino and matter fields at the Gargamelle experiment at CERN~\cite{GargamelleNeutrino:1973jyy}. In 1977, Wolfenstein suggested non-standard lepton flavor violating NC interaction between neutrinos and matter fields as a solution to the solar neutrino anomaly~\cite{Wolfenstein:1977ue}. As is well-known this suggestion was eventually ruled out and replaced by a neutrino mixing scheme. However, in recent years NSI has gained renewed interest as a subdominant effect to be discovered in current and upcoming precision neutrino experiments. The effective  Lagrangian of the NC NSI  between neutrinos and matter fields  can be written as
\begin{align}
    \mathcal{L}_{\rm NSI}^{\rm NC} = \sum_{\alpha,\beta=e,\mu,\tau \atop q=u,d,s} \frac{G_{\rm F}}{\sqrt{2}} \left[ \overline{\nu}_\alpha \gamma^{\mu} \left( 1-\gamma^5 \right) \nu_{\beta} \right] \left[ \overline{q} \gamma_{\mu} \left( \epsilon_{\alpha \beta}^{V q}+\epsilon_{\alpha \beta}^{A q} \gamma^{5} \right) q \right],
    \label{eq:NC-NSI-Lagrangian}
\end{align}
where $G_{\rm F}$ is the Fermi constant. 
The dimensionless parameters $\epsilon_{\alpha \beta}^{Vq}$ and  $\epsilon_{\alpha \beta}^{Aq}$ respectively quantify the vector and axial NSI couplings.
In the limit $\epsilon_{\alpha \beta}^{Vq},\epsilon_{\alpha \beta}^{Aq}\to 0$, 
we recover the SM.

The vector part of the Lagrangian can affect the pattern of neutrino oscillation in matter~\cite{Roulet:1991sm} as well as the Coherent Elastic $\nu$ Nucleus Scattering (CE$\nu$NS)~\cite{Barranco:2005yy}. As a result, it has received wide interest in the literature both from model-building and experimental points of view. The present data already severely constrains the values of $\epsilon^{Vq}_{\alpha \beta}$~\cite{Esteban:2018ppq}. 
By comparison, the axial couplings $\epsilon^{Aq}_{\alpha \beta}$ are less constrained because these couplings cannot affect the neutrino oscillation pattern or CE$\nu$NS which are the main sources of information on $\epsilon^{Vq}_{\alpha \beta}$. The $\epsilon^{Aq}_{\alpha e}$ and $\epsilon^{Aq}_{\alpha \mu}$ have been probed by Deep Inelastic Scattering (DIS) of the $\overline{\nu}_e$ and $\nu_\mu$ beams at the CHARM and NuTeV experiments, respectively~\cite{Davidson:2003ha,Escrihuela:2009up}. The combination $\epsilon^{Au}_{\alpha \beta}-\epsilon^{Ad}_{\alpha \beta}$ is constrained by SNO~\cite{Coloma:2023ixt,Abbaslu:2023vqk}. However, the bounds on $\epsilon^{Aq}_{\tau \tau}$ and $\epsilon^{Aq}_{\tau e}$ are still very weak. In Ref.~\cite{Abbaslu:2023vqk}, we discussed how studying NC DIS at DUNE can improve the bounds on $\epsilon^{Aq}_{\alpha \beta}$. In particular, we demonstrated that $\epsilon^{Aq}_{\tau \tau}$ can be probed for values well below the present bounds by the far detector of DUNE. In Ref.~\cite{Abbaslu:2024hep}, we proposed a toy model leading to $\epsilon^{Au}_{\tau \tau}=\epsilon^{Ad}_{\tau \tau}\sim 1$ with a dark matter candidate as a bonus to be tested by the spin-dependent direct dark matter search experiments. It is thrilling that new physics, involving the great mystery of dark matter, can show up in the measurements of NC interactions of neutrinos with matter fields, ushering in the ``next" SM of particles.

NSI can also affect the resonance and QE interaction of neutrinos with matter. The impact of vector NC NSI on the QE neutrino-nucleus scattering has been discussed in Ref.~\cite{Papoulias:2016edm} (see also Refs.~\cite{Papoulias:2013gha,Papoulias:2015vxa,Papoulias:2015iga}) with special attention to a degeneracy induced by the strangeness form factor.
Recent paper \cite{Ilma:2024lkp} discusses the impact of 
the $\mu\mu$ component of NSI on the QE cross section with special attention to the polarization of the outgoing nucleons. Ref. \cite{Gehrlein:2024vwz} uses the NO$\nu$A data to constrain NSI.

In this work, we shall focus on the impact of axial NSI on QE and resonance interactions. Considering that $\epsilon_{\alpha \beta}^{Aq}$ is less constrained, the impact can be sizable and of relevance to a broadband experiment such as DUNE as well as the medium and low energy on-going and future experiments such as NO$\nu$A, T2K, and T2HK. We shall discuss in detail how uncertainties on the relevant form factors can affect the derivation of information on $\epsilon_{\alpha \beta}^{Aq}$. Moreover, we discuss how by combining different approaches, we can solve degeneracies. This analysis, which focuses on the uncertainties induced by various form factors, can identify the bottlenecks and illuminate the way to improve the theoretical uncertainties in the form factors to uncover new physics. Such an insight can hardly be achieved in the complicated analysis of real or mock data plagued with experimental uncertainties and errors.

The paper is organized as follows. In Sect.~\ref{sec:had-cu}, we present the hadronic current in the presence of NSI. In Sect.~\ref{sec:constraints}, we review the present bounds in the literature on the NSI couplings. We point out an unsolvable degeneracy for which NSI leads to $J_{\rm had}^\mu \to -J_{\rm had}^\mu$. In Sect.~\ref{sec:deg}, we review the methods for deriving the form factors of the QE and resonance scatterings and discuss how NSI shows up in these methods. We suggest combining the different methods to unravel the effects of NSI. In Sect.~\ref{sec:figs}, we demonstrate the impact of NSI on the cross sections of scattering of neutrinos and antineutrinos off the Argon nucleus, taking into account the nuclear effects on scattering. In Sect.~\ref{summary}, we review our results. Appendices~\ref{sect:App-a} and \ref{app-b} review the definition of the form factors for respectively QE and resonance scatterings.

\section{The hadronic currents in the presence of NSI}
\label{sec:had-cu}
Similarly to the standard NC interaction, the effective interaction in Eq.~(\ref{eq:NC-NSI-Lagrangian}) can be written as the product of the leptonic and hadronic currents.
Let us parametrize the total (SM+NSI) four-Fermi neutrino quark NC interactions as
\begin{align} 
    \frac{G_{\rm F}}{\sqrt{2}}\left[\overline{\nu}_\alpha \gamma^\mu (1-\gamma_5)\nu_\beta\right](J_{\rm had}^\mu)_{\alpha \beta}.
    \label{eq:JJ}
\end{align}

In the absence of NSI, the leptonic current is flavor universal and the hadronic current will be proportional to $\delta_{\alpha \beta}$. However, in the presence of NSI, $J_{\rm had}$ can depend on the lepton flavor indices $\alpha$ and $\beta$.
The hadronic NC can be  decomposed as vector and axial components:
\begin{align}
    (J_{\rm had}^\mu)_{\alpha\beta}=(V_{\rm NC}^\mu)_{\alpha\beta}+(A_{\rm NC}^\mu)_{\alpha\beta}\ .
    \label{eq:Jahed}	
\end{align}
Let us rewrite each of these currents as the sum of SM and NSI contributions
\begin{align}
    V_{\rm NC}^\mu=V_{\rm SM}^\mu+V_{\rm NSI}^\mu \quad
    {\rm and} \quad
    A_{\rm NC}^\mu=A_{\rm SM}^\mu +A_{\rm NSI}^\mu
    \notag
\end{align}
where  
\begin{align} 
    (V_{\rm SM}^\mu)_{\alpha\beta}=\left[(1-2\sin^2\theta_W) \overline{Q} \gamma^\mu \frac{\tau^3}{2} Q-\frac{\sin^2\theta_W}{3} \overline{Q}\gamma^\mu Q	+(\frac{2}{3}\sin^2 \theta_W-\frac{1}{2})\overline{s}\gamma^\mu s\right] \delta_{\alpha\beta} 
    \label{eq:VSM}
\end{align}
and
\begin{align} 
    (A_{\rm SM}^\mu)_{\alpha\beta}= \left[-\overline{Q} \gamma^\mu\gamma^5 \frac{\tau^3}{2} Q+\frac{1}{2}\overline{s}\gamma^\mu \gamma^5 s\right] \delta_{\alpha\beta}
    \label{eq:ASM}
\end{align}
in which $Q=(u \ d)^T$ and $\tau^3=\rm{diag} ({1,-1})$ is a  Pauli matrix. Since we are interested only in resonance and QE interaction with nuclei, we have not written the contribution from heavier quarks to the SM current. The NSI currents can be written as 
\begin{align}
    (V^\mu_{\rm NSI})_{\alpha \beta}= \frac{\epsilon^{Vu}_{\alpha \beta} +\epsilon^{Vd}_{\alpha \beta}}{2}\overline{Q}\gamma^\mu Q+ \frac{\epsilon^{Vu}_{\alpha \beta} -\epsilon^{Vd}_{\alpha \beta}}{2}\overline{Q}\gamma^\mu \tau^3 Q+\epsilon^{Vs}_{\alpha \beta} \overline{s}\gamma^\mu s
    \label{eq:VNSI}
\end{align}
and
\begin{align}
    (A^\mu_{\rm NSI})_{\alpha \beta}= \frac{\epsilon^{Au}_{\alpha \beta} +\epsilon^{Ad}_{\alpha \beta}}{2}\overline{Q}\gamma^\mu \gamma^5 Q+ \frac{\epsilon^{Au}_{\alpha \beta} -\epsilon^{Ad}_{\alpha \beta}}{2}\overline{Q}\gamma^\mu  \gamma^5\tau^3 Q+\epsilon^{As}_{\alpha \beta} \overline{s}\gamma^\mu \gamma^5 s . 
    \label{eq:ANSI}
\end{align}
Notice that while $\overline{Q}\gamma^\mu Q$, $\overline{Q}\gamma^\mu\gamma^5 Q$, $\overline{s}\gamma^\mu s$ and $\overline{s}\gamma^\mu \gamma^5 s$ are all isospin scalar (isoscalar), the operators $\overline{Q}\gamma^\mu \tau^3 Q$ and  $\overline{Q}\gamma^\mu\gamma^5 \tau^3 Q$ are isospin vector (isovector).
All these operators can contribute to the amplitudes of QE scatterings as well as to that of resonant isospin 1/2 particle production.  
In the model proposed in Ref.~\cite{Abbaslu:2024hep},  
$\epsilon_{\tau \tau}^{Au}=\epsilon_{\tau \tau}^{Ad}\sim 1$.

\section{Uncharted territory in the parameter space of NSI}
\label{sec:constraints}
There are four main classes of observations and experiments that can probe the
NSIs of the neutrinos with the nuclei. In the following, we enumerate them, pointing out which combination of NSI each experiment is sensitive to. 
\begin{itemize}
    \item {\it Neutrino oscillation experiments:} 
    Neutrino oscillation pattern in matter  is sensitive to $\epsilon_{\alpha \beta}^{V q}\Big|_{\alpha \ne \beta}$ and to $\epsilon_{\alpha \alpha}^{V q}- \epsilon_{\beta \beta}^{V q}$. As a result, solar, atmospheric, and long baseline neutrino oscillation experiments can probe these combinations of NSI \cite{Coloma:2023ixt}.
    \item {\it High energy neutrino beam scattering experiments:} 
    The DIS of neutrinos is sensitive to the following chiral couplings:
    \begin{align}
        \epsilon^{Rq}=\frac{\epsilon^{Vq}+\epsilon^{Aq}}{2} \quad 
        {\rm and} \quad
        \epsilon^{Lq}=\frac{\epsilon^{Vq}-\epsilon^{Aq}}{2}.
        \notag
    \end{align}
    The CHARM experiment with  the $\nu_e$ and $\overline{\nu}_e$  beams was a DIS experiment and was therefore sensitive to the combinations \cite{Davidson:2003ha}
    \begin{align} 
        (g_L^u+\epsilon_{ee}^{Lu})^2+\sum_{\alpha \ne e}|\epsilon_{\alpha e}^{Lu}|^2+ (g_L^d+\epsilon_{ee}^{Ld})^2+\sum_{\alpha \ne e}|\epsilon_{\alpha e}^{Ld}|^2,
        \label{eq:CHARML} 
    \end{align}
    and
    \begin{align} 
        (g_R^u+\epsilon_{ee}^{Ru})^2+\sum_{\alpha \ne e}|\epsilon_{\alpha e}^{Ru}|^2+ (g_R^d+\epsilon_{ee}^{Rd})^2+\sum_{\alpha \ne e}|\epsilon_{\alpha e}^{Rd}|^2,
        \label{eq:CHARMR}
    \end{align}
    in which
    \begin{align}
        g_L^u=\frac{1}{2}-\frac{2}{3} \sin^2 \theta_{\rm W}, \
        g_L^d=-\frac{1}{2}+\frac{1}{3} \sin^2 \theta_{\rm W}, \
        g_R^u=-\frac{2}{3} \sin^2 \theta_{\rm W} \ {\rm and} \
        g_R^d=\frac{1}{3} \sin^2 \theta_{\rm W}.
    \end{align}
    Similarly, the DIS scattering at NuTeV experiment with the $\nu_\mu$ and $\overline{\nu}_\mu$ beams was sensitive to the combinations \cite{Davidson:2003ha}
    \begin{align} 
        (g_L^u+\epsilon_{\mu\mu}^{Lu})^2+\sum_{\alpha \ne \mu}|\epsilon_{\alpha\mu }^{Lu}|^2+ (g_L^d+\epsilon_{\mu\mu}^{Ld})^2+\sum_{\alpha \ne \mu}|\epsilon_{\alpha \mu}^{Ld}|^2,
        \label{eq:NUTEVL} 
    \end{align}
    and
    \begin{align} 
    (g_R^u+\epsilon_{\mu\mu}^{Ru})^2+\sum_{\alpha \ne \mu}|\epsilon_{\alpha \mu}^{Ru}|^2+ (g_R^d+\epsilon_{\mu\mu}^{Rd})^2+\sum_{\alpha \ne\mu}|\epsilon_{\alpha \mu}^{Rd}|^2.
    \label{eq:NUTEVR} 
    \end{align}
    \item {\it The CE$\nu$NS experiment:}
    The coherent elastic scattering of $\nu_\alpha$ off a nucleus with $Z$ protons and $A-Z$  neutrons is sensitive to 
    \begin{align}
        [Z g_p^V+(A-Z)g_n^V+(Z+A)\epsilon_{\alpha\alpha}^{Vu}+(2A-Z)\epsilon_{\alpha\alpha}^{Vd}]^2+\sum_{\alpha \ne \beta} [(Z+A)\epsilon_{\alpha\beta}^{Vu}+(2A-Z)\epsilon_{\alpha\beta}^{Vd}]^2 
        \label{eq:7charge}
    \end{align}
    in which  $g_n^V=-1/2$ and $g_p^V=1/2-2\sin^2 \theta_{\rm W}$.
    \item {\it The SNO experiment:}
    The Deuteron dissociation at SNO was sensitive to  $\epsilon_{\alpha \beta}^{Au}-\epsilon_{\alpha \beta}^{Ad}$. It is noteworthy that SNO NC measurement, along with the solution  $\epsilon_{\alpha \beta}^{Au}\simeq \epsilon_{\alpha \beta}^{Ad} $, has non-trivial solutions such as   $\epsilon_{ee}^{Au}-\epsilon_{ee}^{Ad}\simeq +2$ and/or $\epsilon_{\mu\mu}^{Au}-\epsilon_{\mu\mu}^{Ad}=+2$ and/or  $\epsilon_{\tau\tau}^{Au}-\epsilon_{\tau\tau}^{Ad}=+2$ \cite{Coloma:2023ixt}.\footnote{Notice that our notation and that of Ref.~\cite{Coloma:2023ixt} have a sign difference in the definition of the NSI parameters $\epsilon_{\alpha \beta}$.} There are also solutions for the SNO data with nonzero off-diagonal $\epsilon^{Aq}_{\alpha \beta}$ but these solutions are already ruled out by other considerations such as the bounds from the CHARM experiment \cite{Coloma:2023ixt}.
\end{itemize} 
Let us now discuss possible degeneracies and then review the existing bounds.
As is well-known, the oscillation data alone finds a solution with large $\epsilon_{\mu\mu}^{Vq}-\epsilon_{ee}^{Vq}=\epsilon_{\tau\tau}^{Vq}-\epsilon_{ee}^{Vq}$ known as LMA-Dark solution. However, with a mediator heavier than $\sim 50$ MeV, this solution is ruled out at a high confidence level with scattering experiments. Thus, we shall not consider the LMA-Dark solution in our analysis.
From Eqs.~(\ref{eq:CHARML}, \ref{eq:CHARMR}), we observe a $2^4=16$ fold degeneracy between the SM solution $\epsilon_{ee}^{Lu}=\epsilon_{ee}^{Ru}=\epsilon_{ee}^{Ld}=\epsilon_{ee}^{Rd}=0$ and the non-trivial cases where one or several of the following relations are satisfied:
\begin{align}
    \epsilon^{Lu}_{ee}=-1+\frac{4}{3} \sin^2 \theta_W,\ 
    \epsilon^{Ru}_{ee}=\frac{4}{3} \sin^2 \theta_W,\  
    \epsilon^{Ld}_{ee}=1-\frac{2}{3} \sin^2 \theta_W,\ {\rm and} \
    \epsilon^{Rd}_{ee}=-\frac{2}{3} \sin^2 \theta_W. 
    \label{eq:degenerate}
\end{align}
From Eqs.~(\ref{eq:NUTEVL}, \ref{eq:NUTEVR}), we obtain another $2^4$ fold degeneracy replacing $ee \to \mu \mu$. In order to avoid the constraints from the neutrino oscillation experiments, we can take $\epsilon_{ee}^{Vq}=\epsilon_{\mu \mu}^{Vq}=\epsilon_{\tau \tau}^{Vq}$.

On the other hand, from Eq.~(\ref{eq:7charge}), we find that the CE$\nu$NS results for arbitrary $A$ and $Z$ is degenerate with the SM provided that
\begin{align}
    \epsilon_{\alpha \alpha}^{Vd}=1-\frac{4}{3}\sin^2 \theta_W  \quad 
    {\rm and} \quad
    \epsilon_{\alpha \alpha}^{Vu}=-1+\frac{8}{3}\sin^2 \theta_W. 
    \label{eq:7deg}
\end{align}
Indeed for $A\simeq 2Z$, as long as $\epsilon^{Vu}_{\alpha \alpha}+\epsilon^{Vd}_{\alpha \alpha}=\frac{4}{3}\sin^2 \theta_W$ the degeneracy holds valid. This explains the solution presented in the second column of Tab.~3 of Ref.~\cite{Coloma:2023ixt}. Taking $\epsilon^{A/V q}_{\alpha \beta} \propto \delta_{\alpha \beta}$ (or equivalently $\epsilon^{L/R q}_{\alpha \beta} \propto \delta_{\alpha \beta}$), we find that satisfying all the relations in Eq.~(\ref{eq:degenerate}) automatically guarantees the relations in Eq.~(\ref{eq:7deg}). In other words, there is a non-trivial solution for NSI that avoids bounds from oscillation data, from CE$\nu$NS as well as from DIS experiments, NuTeV and CHARM. Moreover, relations in Eq.~(\ref{eq:degenerate}) imply $\epsilon_{\alpha \alpha}^{Au}-\epsilon_{\alpha \alpha}^{Ad}=+2$ which is one of the non-trivial solutions of SNO NC measurement presented in Tab.~4 of Ref.~\cite{Coloma:2023ixt}. 

It is straightforward to confirm that going from the $\epsilon_{\alpha \beta}=0$ solution to $\epsilon_{ee}=\epsilon_{\mu\mu}=\epsilon_{\tau\tau}$ with relations in Eq.~(\ref{eq:degenerate}),  $J_{\rm had}^\mu$ transforms to its opposite, up to the contribution from the strange currents $\overline{s} \gamma^\mu s$ and $\overline{s} \gamma^\mu \gamma^5 s$. 
Notice however that if, in addition to satisfying relations in Eq.~(\ref{eq:degenerate}), we take $\epsilon_{\alpha \beta}^{A/Vs}=\epsilon_{\alpha \beta}^{A/Vd}$, then $J_{\rm had}^\mu\to -J_{\rm had}^\mu$ so with measuring the scattering cross sections, the two solutions cannot be distinguished.

Apart from this degenerate solution which seems to be difficult to disentangle, the only surviving solution for $\epsilon^{Vq}$ is the solution around zero (the SM limit) with  $\epsilon^{Vu},\epsilon^{Vd}\stackrel{<}{\sim} 0.02$ (for more details see the second column of Tab.~3 of \cite{Coloma:2023ixt}). 
From NuTeV Ref.~\cite{Davidson:2003ha}, there are strong bounds\footnote{As discussed in Ref.~\cite{Abbaslu:2023vqk}, we should, however, take the NuTeV bounds with a grain of salt because the derivation of the bound is based on a relation for the ratio of NC to Charged Current (CC) cross sections which neglects the subdominant effects from the sea $s$-quarks of the nucleons. We estimate the effect of this overlooked contribution to be of the order of $\mathcal{O}(3\%)$ so the derived bounds cannot be stronger than few$\times 0.01.$} on $\epsilon_{\mu \mu(\tau)}^{A u/d}$:
\begin{align} 
    \epsilon^{A u/d}_{\mu\mu}<0.01 \quad
    {\rm and} \quad
    \epsilon^{A u/d}_{\mu\tau}<0.005.
\end{align}
Comparing Eqs.~(\ref{eq:ANSI}, \ref{eq:VNSI}) with Eqs.~(\ref{eq:ASM}, \ref{eq:VSM}), we observe that for the contributions from $\epsilon^{Au}_{\mu \mu}-\epsilon^{Ad}_{\mu \mu}$ and from $\epsilon^{Vu}_{\alpha\alpha}-\epsilon^{Vd}_{\alpha\alpha}$ to the cross sections of the relevant processes will be negligible and less than respectively $(\epsilon^{Au}_{\mu \mu}-\epsilon^{Ad}_{\mu \mu})/(1/2)=4 \%$  and $( \epsilon_{\alpha \alpha}^{Vu}-\epsilon_{\alpha \alpha}^{Vd})/[(1-2\sin^2\theta_W)/2]=16 \%$ of the SM contributions.
Any contribution from lepton flavor violating components $\epsilon_{\alpha \beta}$ will be proportional to the square of these NSI couplings so the contribution from  $\epsilon_{\mu \tau}^{Au}-\epsilon_{\mu \tau}^{Ad}$ and $(\epsilon_{\alpha \beta}^{Vu}-\epsilon_{\alpha \beta}^{Vd})|_{\alpha \ne \beta}$ will be further suppressed by respectively factors of $(\epsilon_{\mu \tau}^{Au}-\epsilon_{\mu \tau}^{Ad})^2$ and $(\epsilon_{\alpha \beta} ^{Vu}-\epsilon_{\alpha \beta} ^{Vd})^2/(1-2\sin^2\theta_W)^2$  and will be therefore negligible.
The bounds on $\epsilon_{\tau \tau}^{Au}+\epsilon_{\tau \tau}^{Ad}$,  $\epsilon_{\tau e}^{Au}+\epsilon_{\tau e}^{Ad}$, $\epsilon_{ee}^{Au}+\epsilon_{ee}^{Ad}$ as well as on $\epsilon_{\alpha \beta}^{Vs}$ and on $\epsilon_{\alpha \beta}^{As}$ are weak. Indeed, they can be of the order of 1 so we will discuss their impact on QE and resonance interaction. 
Recent paper \cite{Gehrlein:2024vwz} uses the NO$\nu$A data to constrain $\epsilon^{An}=\epsilon^{Ap}$ with the result $-0.41 <\epsilon^{An}_{\tau \tau}=\epsilon^{Ap}_{\tau \tau}<0.41$ and $-0.40 <\epsilon^{An}_{e \tau}=\epsilon^{Ap}_{e \tau}<0.40$. Using the formulas in \cite{Abbaslu:2023vqk,Cirelli:2013ufw,Belanger:2013oya}, $\epsilon^{An}=\epsilon^{Ap}=0.41\epsilon^{Au}=0.41\epsilon^{Ad}$ which means the bounds are not stronger than $\mathcal{O}(1)$.
The bounds from SNO on $\epsilon^{Au}_{\alpha \beta}-\epsilon^{Ad}_{\alpha \beta}$ are as follows \cite{Coloma:2023ixt}
\begin{align}
    -0.13<\epsilon_{ee}^{Au}-\epsilon_{ee}^{Ad}<0.19 \quad
    {\rm and} \quad 
    -0.15<\epsilon_{\tau\tau}^{Au}-\epsilon_{\tau\tau}^{Ad}<0.2,
    \label{eq:eAdiag}
\end{align}
and
\begin{align}
    -0.12<\epsilon_{e\mu}^{Au}-\epsilon_{e\mu}^{Ad}<0.16 \quad 
    {\rm and} \quad
    -0.1<\epsilon_{e\tau}^{Au}-\epsilon_{e\tau}^{Ad}<0.13.
\end{align}
The contribution from the off-diagonal elements will be suppressed by their square and will be therefore negligible. However, we shall discuss the possible  effects of $\epsilon^{Au}_{ee}-\epsilon^{Ad}_{ee}$ and $\epsilon^{Au}_{\tau\tau}-\epsilon^{Ad}_{\tau\tau}$.

\section{Degeneracies between form factors and NSI couplings\label{sec:deg}}
To compute the amplitude of QE or resonance scattering ({\it i.e.,} $\mathcal{M}(\nu+N(p)\to \nu+N(p'))$ or $\mathcal{M}(\nu+N(p)\to \nu+R(p'))$ where $R\in \{N',\Delta,...\}$), 
we should sandwich  the quark current between the initial and final hadron states: 
\begin{align}
    \langle N(p')|J_{\rm had}^\mu|N(p)\rangle \quad
    {\rm and} \quad
    \langle R(p')|J_{\rm had}^\mu|N(p)\rangle. 
\end{align}
The matrix elements are given by the relevant form factors which are functions of $Q^2=-(p-p')^2$. The form factors cannot be perturbatively computed but through techniques such as lattice QCD \cite{Alexandrou:2018sjm,Alexandrou:2021wzv,Alexandrou:2022yrr} or sum rules \cite{Azizi:2015fqa}, their values can be predicted with a limited precision. On the other hand, various scattering experiments are sensitive to certain different combinations of form factors. As we shall see in this section, in the presence of NSI, the extraction of the values of the form factors from the neutrino scattering experiments can be subject to new degeneracies. In this section, we discuss the uncertainties and degeneracies and suggest approaches that can in principle solve them.
In Sect.~\ref{sec:quasi}, we focus on the QE interactions. In Sect.~\ref{sec:resonances}, we discuss the resonant scatterings.

\subsection{Quasi-Elastic scatterings}
\label{sec:quasi}
As is well-known the cross section of the QE scattering is given by 
\begin{align}
    &\frac{d\sigma^{\text{QE}}_{\text{NC}}\big( \overset{(-)}{\nu}_{\alpha} + N \to \overset{(-)}{\nu}_{\beta} + N \big)}{dQ^2}= \notag\\
    &\frac{G_F^2 Q^2}{2 \pi E^{2}_{\nu}} \left[ A^N(Q^2) \pm B^N(Q^2) \left( \frac{4 E_{\nu}}{M_N} - \frac{Q^2}{M_N^2} \right) + C^N(Q^2) \left( \frac{4 E_{\nu}}{M_N} - \frac{Q^2}{M_N^2} \right)^2 \right],
    \label{eq:diff-sig}
\end{align}
where the plus (minus) sign is for the neutrino (antineutrino) scattering. The functions $A(Q^2)$, $B(Q^2)$, and $C(Q^2)$ are defined as follows:
\begin{align}
    &A^N(Q^2) = \notag \\&\frac{1}{4} \left[ 
    \left( \tilde{F}_{A}^{N} \right)^2 \left(1 + \frac{Q^2}{4M_N^2}\right) 
    - \left(\left( \tilde{F}_{1}^{N} \right)^2 - \frac{Q^2}{4M_N^2} \left( \tilde{F}_{2}^{N} \right)^2 \right)
    \left(1 - \frac{Q^2}{4M_N^2}\right)+ \left(\frac{Q^2}{M_N^2}\right) \tilde{F}_{1}^{N} \tilde{F}_{2}^{N}\right],
\end{align}
\begin{align}
    B^N(Q^2) &= - \frac{1}{4} \tilde{F}_{A}^{N} \left( \tilde{F}_{1}^{N} + \tilde{F}_{2}^{N} \right), \\
    C^N(Q^2) &= \frac{M_N^2}{16 Q^2} \left[
    \left( \tilde{F}_{A}^{N} \right)^2 + \left( \tilde{F}_{1}^{N} \right)^2 + \left(\frac{Q^2}{4M_N^2}\right) \left( \tilde{F}_{2}^{N} \right)^2\right].
\end{align}

In the appendix~\ref{sect:App-a}, we review the definitions of $\tilde{F}$ within the SM and discuss their energy dependence.
From Eq.~(\ref{eq:diff-sig}), we observe that the difference between neutrino and antineutrino cross sections yields $B^N$ or equivalently, $\tilde{F}_A^N(\tilde{F}_1^N+\tilde{F}_2^N)$. Moreover, the  $E_\nu$ dependence can determine $A^N$, $B^N$ and $C^N$ which are combinations of $\tilde{F}_A^N$, $\tilde{F}_1^N$ and $\tilde{F}_2^N$.
Thus, by studying the scattering of neutrino and antineutrino beams off the nucleons, $\tilde{F}_1^N$, $\tilde{F}_A^N$ and $\tilde{F}_2^N$ can in principle be extracted.

From Eqs.~(\ref{eq:Jahed}-\ref{eq:ANSI}), we observe that the effects of NSI can be accounted for by redefining $\tilde{F}$ as

\begin{align}
    (\tilde{F}^p_i)_{\alpha \beta} =&\, \left( \frac{\delta_{\alpha \beta}}{2} - 2 \delta_{\alpha \beta} \sin^2 \theta_W + 2 \epsilon^{V u}_{\alpha \beta} + \epsilon^{V d}_{\alpha \beta} \right) F^p_i + \left( -\frac{\delta_{\alpha \beta}}{2} + \epsilon^{V u}_{\alpha \beta} + 2 \epsilon^{V d}_{\alpha \beta} \right) F^n_i \notag \\
    &+ \left( -\frac{\delta_{\alpha \beta}}{2} + \epsilon^{V u}_{\alpha \beta} + \epsilon^{V d}_{\alpha \beta} + \epsilon^{V s}_{\alpha \beta} \right) F^s_i,\\
    (\tilde{F}^n_i)_{\alpha \beta} =&\, \left( \frac{\delta_{\alpha \beta}}{2} - 2 \delta_{\alpha \beta} \sin^2 \theta_W + 2 \epsilon^{V u}_{\alpha \beta} + \epsilon^{V d}_{\alpha \beta} \right) F^n_i + \left( -\frac{\delta_{\alpha \beta}}{2} + \epsilon^{V u}_{\alpha \beta} + 2 \epsilon^{V d}_{\alpha \beta} \right) F^p_i\notag \\
    &+ \left( -\frac{\delta_{\alpha \beta}}{2} + \epsilon^{V u}_{\alpha \beta} + \epsilon^{V d}_{\alpha \beta} + \epsilon^{V s}_{\alpha \beta} \right) F^s_i,
\end{align}

\begin{align}
(\tilde{F}^p_A)_{\alpha \beta} &= \left( -\frac{\delta_{\alpha \beta}}{2} + \frac{\epsilon^{A u}_{\alpha \beta} - \epsilon^{A d}_{\alpha \beta}}{2} \right) F_A + \frac{3}{2} (\epsilon^{A u}_{\alpha \beta} + \epsilon^{A d}_{\alpha \beta}) F^{(8)}_A + \left( \frac{\delta_{\alpha \beta}}{2} + \epsilon^{A u}_{\alpha \beta} + \epsilon^{A d}_{\alpha \beta} + \epsilon^{A s}_{\alpha \beta} \right) F^s_A, \label{eq:FAp}\\
(\tilde{F}^n_A)_{\alpha \beta} &= \left( \frac{\delta_{\alpha \beta}}{2} - \frac{\epsilon^{A u}_{\alpha \beta} - \epsilon^{A d}_{\alpha \beta}}{2} \right) F_A + \frac{3}{2} (\epsilon^{A u}_{\alpha \beta} + \epsilon^{A d}_{\alpha \beta}) F^{(8)}_A + \left( \frac{\delta_{\alpha \beta}}{2} + \epsilon^{A u}_{\alpha \beta} + \epsilon^{A d}_{\alpha \beta} + \epsilon^{A s}_{\alpha \beta} \right) F^s_A.
\label{eq:FAn}
\end{align}

The definitions of $F_i^p$, $F_i^n$, $F_A$ and $F_A^s$ can be found in appendix~\ref{sect:App-a}.
Notice, however, that in the presence of axial NSI a new form factor enters the cross section formula defined as 
\begin{align}
   \langle N(p^{\prime})| \overline{q}\gamma^{\mu}\gamma^5 \frac{\lambda_8 }{\sqrt{3}}q |N(p)\rangle=  F^{(8)}_A(Q^2)\overline{u}_N(p^{\prime})\gamma^{\mu}\gamma^5u_{N}(p)+\frac{F_P^{(8)}(Q^2)}{M_N}\overline{u}_N(p')q^\mu\gamma^5 u_N(p),   
\end{align}
where $\lambda_8$ is the 8th Gell-Mann matrix: ${\rm diag}(1/\sqrt{3},1/\sqrt{3},-2/\sqrt{3})$.
In the following, we first discuss  $\tilde{F}_i^N$ and vector NSI. We then focus on 
$\tilde{F}_A^N$ and axial NSI.

{\it $\tilde{F}^N_i$ and vector NSI:} As explained in appendix~\ref{sect:App-a}, the form factors $F_1^N$ and $F_2^N$ are directly related to the electric charge and magnetic dipole of the nucleon, $N$.
Not surprisingly, $F_1^p$ and $F_2^p$  have been extracted as a function of $Q^2$ with a remarkable precision from the electron proton scattering experiments \cite{A1:2013fsc,Punjabi:2015bba}. As long as NSI does not involve the electron, these derivations are free from neutrino NSI ($\epsilon_{\alpha \beta}^{V/A q}$) contamination.
As discussed above, $\tilde{F}^N_i$ can also be derived from studying the scattering of neutrino and antineutrino beams.
Within the SM, the knowledge of $\tilde{F}^N_i$ and $F_i^N$ can yield $F_i^s$. Indeed the $\nu_\mu$ scattering data from MiniBooNE has been used to constrain $F_i^s$ \cite{Pate:2024acz}, showing that $-0.09<F_1^s(Q^2)<0.1$ for $Q^2<1$ GeV$^2$ and tighter bounds for smaller $Q^2$.  On the other hand, $F_i^s(Q^2)$ are predicted by lattice QCD \cite{Alexandrou:2019olr} with the result $F_1^s (0)=0$, $F_1^s (Q^2)<0.004$ and $F_2^s(0)=-0.017$. Moreover, the global analysis of parity violating electron-proton scattering experiments yields $F_i^s$ with a value consistent with the lattice QCD results \cite{Liu:2007yi}. While the computation of lattice QCD, as well as the derivation from the parity violating electron-proton scattering experiments, are not affected by NSI, according to Eq.~(\ref{eq:FAp}) the derivation of $F_i^s$ from the $\nu_\mu$ scattering off protons in the presence of NSI should be interpreted as $(2 \epsilon_{\mu\mu}^{Vu}+\epsilon_{\mu\mu}^{Vd})F_i^p+(\epsilon_{\mu\mu}^{Vu}+2\epsilon_{\mu\mu}^{Vd})F_i^n+(-1/2+\epsilon_{\mu\mu}^{Vs})F_i^s$ where we have used $\epsilon_{\mu\mu}^{Vu},\epsilon_{\mu\mu}^{Vd}\ll 1/2$.
Despite the strong bounds on $\epsilon_{\mu\mu}^{Vq}$,  $F_i^p \epsilon_{\mu\mu}^{Vu}$ or  $F_i^p \epsilon_{\mu\mu}^{Vd}$ can still be larger than $F_i^s/2$. If, contrary to the expectations, future measurements establish $F_1^s(Q^2)\sim 0.01$, it should be interpreted as nonzero $\epsilon_{\mu\mu}^{Vu}$ and/or $\epsilon_{\mu\mu}^{Vd}$. Notice that $\tilde{F}^p_1(0)$ and $\tilde{F}^n_1(0)$
respectively give $2\epsilon_{\mu\mu}^{Vu}+\epsilon_{\mu\mu}^{Vd}$ and 
$\epsilon_{\mu\mu}^{Vu}+2\epsilon_{\mu\mu}^{Vd}$ (see Eqs.~(\ref{eq:FAp}, \ref{eq:FAn}). Hence, their combination gives $\epsilon_{\mu\mu}^{Vu}$ and $\epsilon_{\mu\mu}^{Vd}$.
Then, $\tilde{F}^p_2$ would give $F_2^s(-1/2+\epsilon_{\mu\mu}^{Vs})$. Inserting the value of $F_2^s$ from lattice QCD prediction, $\epsilon_{\mu\mu}^{Vs}$ can be extracted. In theory, the knowledge of $\tilde{F}_2^n$ as well as that of the $Q^2$-dependence of $\tilde{F}_i^N$ can over-constrain the unknowns. In reality, since various experimental errors induce sizable error bars in the derivation of $\tilde{F}^N_1(0)$ and $\tilde{F}^p_2(0)$, the alternative information from $\tilde{F}_2^n$ can reduce the error bars.

Notice that in the above discussion, we did not include the off-diagonal NSI. Since the determination of the flavor of the final neutrino is not feasible, in the presence of off-diagonal $\epsilon_{\alpha \beta}^{Vq}$, we should sum over the cross sections of all final flavors of $\nu_\mu+N\to \nu_\alpha +N$. For the flavor violating case, the contribution from NSI to the cross section is suppressed by quadratic terms in $\epsilon_{\alpha \beta}^{Vq}$ and $F_i^s \epsilon^{Vs}_{\alpha \beta}$. Thus, the contribution can be neglected.

{\it $\tilde{F}^N_A$ and axial NSI:} We first review the alternative methods for the $F_A$ and $F_A^s$ derivation within the SM. We then discuss how  $F_A^{(8)}$ can be obtained. In the end, we discuss how NSI can show up.

Let us start the discussion with $F_A$. As discussed in appendix~\ref{sect:App-a}, $F_A$ is related to $\langle N|\overline{Q} \gamma^\mu \gamma^5 \tau^3 Q|N\rangle$. The isospin symmetry implies $$\langle N(p')|  \overline{Q} \gamma^\mu \gamma^5 \tau^3 Q| N(p)\rangle=\langle N(p')|  \overline{Q} \gamma^\mu \gamma^5 \tau^{1} Q| N(p)\rangle=\langle N(p')|  \overline{Q} \gamma^\mu \gamma^5 \tau^{2} Q| N(p)\rangle.$$ Since the CC interaction of nucleons is given by the latter two matrix elements, the value of $F_A$ can therefore be derived by studying the CC interaction of leptons with nucleons. The value at $Q^2=0$, $F_A(0)=g_A$, which is called weak axial vector coupling constant, can be derived from the $\beta$-decay experiments. Various collaborations report the value of $g_A$ with high precision. For example, according to Ref.~\cite{Markisch:2018ndu},  $\lvert g_A\lvert=1.27641(55)$.  However, while different derivations agree on the first two digits, there is a discrepancy of $\sim 0.01$ between them.  Anyway, to relate $g_A$ extracted from $\beta$-decay to $g_A$ appearing in the NC interaction of nucleons, we should use isospin symmetry which is accurate only up to 1\%. Thus, the mentioned discrepancy is irrelevant to our discussion. 
At low $Q^2$ ({\it i.e.,} $Q^2<0.6$ GeV$^2$) $F_A(Q^2)$ is usually parametrized as a dipole: 
\begin{align}
    F_A(Q^2)=\frac{g_A}{(1+Q^2/M_A^2)^{2}}.
    \label{eq:FA-}
\end{align}
The extraction of the value of $M_A$, dubbed as axial mass,  is subject to uncertainties and its central fit depends on the assumptions on the $Q^2$ dependence of vector form factors. According to Ref.~\cite{Kuzmin:2007kr}, $M_A=0.999\pm 0.011$. The derivation of $F_A(Q^2)$ from CC interactions is free from degeneracies induced by the NC NSI.
 
Similarly to the case of $F_A(Q^2)$, for low $Q^2$,   $F_A^s(Q^2)$ can be parameterized as a dipole form:
\begin{align}
	F_A^s=\frac{g_A^s}{\left[1+Q^2/(M_A^s)^2\right]^2}\ .
\end{align} 
In principle, $M_A^s$ can be different from $M_A$
but within the present experimental uncertainties, we can set $M_A^s=M_A$. 
For $Q^2>0.5$~GeV$^2$, more elaborate parametrization  should be adopted both for $F_A$ and $F_A^s$ (see Refs.~\cite{Bradford:2006yz}, \cite{Pate:2024acz}). 

There are four approaches to determine $g_A^s$:
\begin{itemize} 
    \item \textit{Lattice QCD}: 
    This determination is independent of NSI and yields {a value of $g_{A}^{s}=-0.044\pm 0.008$ \cite{Alexandrou:2022yrr,Alexandrou:2021wzv,Alexandrou:2019brg}.}
    \item \textit{Measuring parity violating asymmetries in scattering of polarized charge lepton beam off nuclei}\cite{COMPASS:2006mhr,HERMES:2006jyl,EuropeanMuon:1989yki,Alberico:2001sd}: 
    The interference between the NC and electromagnetic contributions leads to parity violation in the scattering of polarized electrons off hadrons. This asymmetry can be used to extract information on the strangeness form factors \cite{Liu:2007yi, Musolf:1993tb, Pate:2024acz}. If NSI only involves neutrinos of certain flavors and the electrons have no NSI this determination will not be sensitive to NSI. The values of $g_A^s$ determined with this method are in good agreement with the values determined by the lattice QCD methods.
    \item \textit{Determining $g_A^s$ from QE scattering of neutrinos off nuclei}:
    Setting all NSI parameters equal to zero, Ref.~\cite{KamLAND:2022ptk} uses the measurement of QE NC interaction of atmospheric neutrinos in KamLAND to determine $g_A^s=-0.14^{+0.25}_{-0.25}$. As we shall discuss, in the presence of NSI, this derivation should be revisited.
    \item \textit{Determining $g_A^s$ from hyperon beta decay}: 
    The global $SU(3)$ flavor symmetry of $T=(u \ d \ s)$ implies a relation between the matrix elements of  $ \overline{T} \gamma^\mu \gamma^5 \lambda_{4,5}T$ and $ \overline{T} \gamma^\mu \gamma^5 \lambda_{8}T$  where $\lambda_i$ are $3\times 3$ Gell-Mann matrices associated to the adjoint representation of  $SU(3)$. The former operators with $\lambda_{4,5}$ convert the $s$-quark to the $u$-quark which can lead to hyperon decay. As a result, the relevant form factor can be extracted by hyperon beta decay. Then, the $SU(3)$ symmetry gives $F_A^{(8)}=(F_A^{(0)}-2 F_A^s)/\sqrt{3}$. We shall not however rely on this approach because the $SU(3)$ symmetry is severely broken. Ref.~\cite{Bass:2009ed} uses this method to extract the form factors.
\end{itemize}

Let us now discuss $F_A^{(8)}(Q^2)$. This form factor is not relevant for the NC interaction within the SM; however, for $\epsilon^{Au}_{\alpha \beta}+ \epsilon^{Ad}_{\alpha \beta} \ne 0$, the NC cross section depends on this form factor. As mentioned above, the global $SU(3)$ symmetry of $(u \ d \ s)$ relates this form factor to the beta decay of hyperon but this $SU(3)$ symmetry is broken. Fortunately, $F_A^{(8)}$ is computed by the lattice QCD \cite{Alexandrou:2021wzv}. We shall use the lattice QCD prediction for $F_A^{(8)}$ which is again free from the degeneracies with the NSI parameters.
 Like the previous axial form factor, for low $Q^2$,	
we can write \cite{Alexandrou:2021wzv}
\begin{align}
    F_A^{(8)}(Q^2)=\frac{g_A^{(8)}}{\left[1+Q^2/(M_A^{(8)})^2\right]^2},
\end{align} 
where the lattice QCD predicts $g_A^{(8)}=0.53 \pm 0.022$ and $M_A^{(8)}=1.154(101)$ \cite{Alexandrou:2021wzv}.

As discussed above by studying the energy dependence of the neutrino current scattering cross section, $\tilde{F}_A^N$ can be derived. Within SM, this measurement combined with $F_A$ from beta decay can determine $F_A^s$. The basis of the $F_A^s$ determination from the atmospheric neutrinos at KamLAND \cite{KamLAND:2022ptk} is this concept. In the presence of NSI, such a derivation should be revisited. Even $(\epsilon_{\mu\mu}^{Au}-\epsilon_{\mu\mu}^{Ad})/2$ as small as 0.02 (within the present bounds) can have a contribution to $\tilde{F}^N_A$ comparable to that from $F_A^s$. More importantly, $(\epsilon_{\tau\tau}^{Au}+\epsilon_{\tau\tau}^{Ad})/2$ which can be $\mathcal{O}(1)$ can dramatically  change the cross section of $\nu_\tau+N\to \nu_\tau+N $. Thus, in the far detectors of long baseline experiments where the  $\nu_\tau$ component of the beam is large, the cross section can significantly deviate from the SM prediction.

If the $Q^2$ dependence of $F_A$, $F_A^{(8)}$ and $F_A^s$  significantly deviated from each other, the $Q^2$ dependence could yield their coefficients and hence $\epsilon^{Au}$, $\epsilon^{Ad}$ and $\epsilon^{As}$, separately. But as we saw above, the axial mass parameters associated to all of them are very close to each other (if not equal). Another tool to investigate NSI is flavor. While $F_A^s$ is a property of nucleon and is independent of the flavor of neutrinos that scatter off the nucleon,  $\epsilon_{\alpha \beta}^{Aq}$ depends on the flavor. Thus, by comparing the derivation of $\tilde{F}_A^n$ at the far and near detectors of long baseline experiments, one can obtain information on $\epsilon_{\alpha \beta}^{Aq}$ and its flavor structure. Another approach to utilize different flavors is studying the energy dependence of the NC event rates at the far detector as $P(\nu_\alpha \to \nu_\beta)$ depends on the energy of the neutrino.

For $E_\nu < 1$~GeV, the oscillation length of $\nu_\mu \to \nu_\tau$ is shorter than 350 km, corresponding to a chord with a zenith angle of 91.5$^\circ$. As a result, for the KamLAND atmospheric neutrino sample with $E_\nu<$GeV which is used in \cite{KamLAND:2022ptk} to extract the value of $F_A^s$, the majority of the neutrino flux coming from below were a mixture of $\nu_\mu$ and $\nu_\tau$ plus a $\nu_e$ component at lower energy. Considering that about half of the neutrinos came from the directions with zenith angle larger than $92^\circ$ for which $\langle P(\nu_\mu \to \nu_\tau)\rangle/\langle P(\nu_\mu \to \nu_\mu)\rangle\simeq 2s_{23}^2c_{23}^2/(c_{23}^4+s_{23}^4) \simeq 1$, we expect about 1/4 of the atmospheric neutrinos studied in Ref.~\cite{KamLAND:2022ptk} to be $\nu_\tau$. In the absence of NSI, Ref.~\cite{KamLAND:2022ptk} derives $F_A^s$ with an accuracy of $\pm 0.25$. Assuming that all the uncertainty was of statistical origin, with $1/4$ of the data the uncertainty would be $\pm 0.5$.
From Eqs.~(\ref{eq:FAp}, \ref{eq:FAn}), we observe that an uncertainty of 0.5 in the determination of $F_A^s$ is roughly equivalent to a bound of about 0.3 on $|\epsilon_{\tau \tau}^{Vu}+\epsilon_{\tau \tau}^{Vd}|$ which is already stronger than the present bound. This estimation of the bound is of course an over-simplification. We expect that with a thorough analysis of the KamLAND atmospheric data, a more stringent bound can be set because most likely a good fraction of the uncertainty is due to systematic errors. That is, by dividing the number of events by four, the deterioration of the sensitivity will be less severe.  A complete analysis of the KamLAND atmospheric data is beyond the scope of the present paper. 
The upcoming more accurate experiments such as JUNO combined with the prediction of $F_A^s$ from lattice QCD can help to set significant bounds on $\epsilon^{Vu}_{\tau \tau}+\epsilon^{Vd}_{\tau \tau}$.
 
\subsection{Resonances \label{sec:resonances}} For energy range $1~{\rm GeV}<E_\nu<4~{\rm GeV}$, the resonance cross section dominates both over QE and DIS. Within the SM, the dominant and therefore best-studied NC resonance is $\nu+N \to \nu+\Delta$, where $\Delta$ has a mass of 1232 MeV, isospin of 3/2, and spin of 3/2.
As shown in Refs.~\cite{Leitner:2006sp,Leitner:2008ue}, the cross section of $\nu+N\to \nu+\Delta$ is described by a special set of form factors denoted by $\tilde{C}_i^A$ and  $\tilde{C}_i^N$.

The resonances with isospin different from $1/2$ cannot receive any contribution from the isospin invariant operators so the $\Delta$ resonance will not be sensitive to $\epsilon^{Au}+\epsilon^{Ad}$,  $\epsilon^{Vu}+\epsilon^{Vd}$, $\epsilon^{Vs}$ or $\epsilon^{As}$.
The effects of $\epsilon^{Au}-\epsilon^{Ad}$ and  $\epsilon^{Vu}-\epsilon^{Vd}$ on the cross section of the process $\nu+N\to \nu+\Delta$ can be treated by the following replacements
\begin{align} \label{CAiCORRECTED}
    \tilde{C}_i^A \delta_{\alpha \beta} \to \tilde{C}_i^A (\delta_{\alpha \beta}-\epsilon^{Au}_{\alpha \beta}+\epsilon^{Ad}_{\alpha \beta}) 
\end{align}
and
\begin{align}
    \tilde{C}_i^V \delta_{\alpha \beta} =  -(1-2\sin^2\theta_W){C}_i^N \delta_{\alpha \beta}\to  -\left[(1-2\sin^2\theta_W) \delta_{\alpha \beta} + \epsilon^{Vu}_{\alpha \beta}-\epsilon^{Vd}_{\alpha \beta}\right]{C}_i^N.
\end{align}
The definitions of $\tilde{C}^A_i$ and $C_i^N$ are shown in Refs.~\cite{Lalakulich:2005cs,Leitner:2006sp,Leitner:2008ue} (see also appendix~ \ref{app-b}).
As discussed before, $\epsilon_{\alpha \beta}^{Vu}-\epsilon_{\alpha \beta}^{Vd}$ are already strongly constrained so the effect will be negligible. 

After the $\Delta$ resonance the next three important resonances have isospin $1/2$ so they can receive  contributions from $\epsilon^{Au}+\epsilon^{Ad}$ and $\epsilon^{Vu}+\epsilon^{Vd}$, too. However, the cross sections of these resonances are small. We therefore do not discuss them further.

\section{Cross sections in the presence of axial NSI}
\label{sec:figs}
In the previous sections, we saw that for certain NSI, we expect a sizable impact on the cross sections of (anti)neutrino nucleus NC QE
and resonance scatterings. In this section, we study the cross sections in the presence of NSI and compare their effect with the uncertainties induced by less known form factors. To compute the cross sections, we use GiBBU 2023 as an event generator that accounts for nuclear effects \cite{Buss:2011mx}, implementing the necessary changes to take into account the NC NSI effects. For the illustrative purposes, we omit the contribution from two particles-two holes (2p-2h) excitations \cite{Delorme:1985ps,Marteau:1999kt,Bleve:2000hc}. Using GiBUU, these excitations can be readily included \cite{Lalakulich:2012ac}.
With the off-diagonal NSI, the scattering can be lepton flavor violating: $\stackrel{(-)}{\nu}_\alpha +N \to \stackrel{(-)}{\nu}_\beta +N$
and $\stackrel{(-)}{\nu}_\alpha+{N} \to \stackrel{(-)}{\nu}_\beta +\Delta$ with $\alpha\neq \beta$. Since the detectors do not determine the flavor of the final neutrino at NC scattering, we sum over the flavor of the final neutrino when computing the NC scattering cross section.
Because of the neutrino oscillation, the incoming beam at the far detector of long baseline experiments will be a coherent combination of all three flavors rather than a pure flavor state. As shown in Ref.~\cite{Abbaslu:2023vqk}, the cross section can be easily computed for such mixed flux using a suitable change of flavor basis. For demonstrative reasons, we carry out our computation for a pure flavor state. For the target, we consider the Argon nucleus.

Let us start with the quasi-elastic scattering. In the end of this section, we will discuss the resonance scattering.
Tab.~\ref{tab:QE} shows the input values of the form factor parameters for the $\stackrel{(-)}{\nu}$-$N$ NC QE scattering and their uncertainties, along with their determination methods and references. The uncertainties of $F_i^N$ are negligible.
For the dependence $Q^2$ of these form factors, we use the BBBA2007 parametrization method \cite{Bodek:2007ym} which is the default of the GiBUU event generator. The effects of $F_i^s(0)$ are negligible and can be set to zero as GiBUU does by default \cite{Leitner:2006sp}. As we discussed in sect. \ref{sec:quasi}, there are several
alternative methods to determine the values of $g_A^s$. The conservative uncertainty that is shown in Tab. \ref{tab:QE} covers all derivations.
\begin{table}[h]
    \caption{The form factor parameters for $\stackrel{(-)}{\nu}-N$ NC QE scattering\label{tab:QE}}
    \begin{ruledtabular}
        \begin{tabular}{c c c c}
            Parameter & Value & Method & Reference\\
            \colrule
            $M_V$ & $0.843\ \rm GeV$ & $\nu-N$ CC QE scattering & \cite{Baker:1981su}\\
            $M_A^{(8)}$ & $1.154 \pm 0.101\ \rm GeV$ & lattice QCD & \cite{Alexandrou:2021wzv}\\
            $M_A$ & $0.999\pm 0.011\ \rm GeV$ & $\nu-N$ CC QE scattering & \cite{Kuzmin:2007kr}\\
            \colrule
            $g_A$ & $1.2695$ & $\beta$-decay & \cite{ParticleDataGroup:2008zun}\\
            $g_A^{(8)}$ & $0.530 \pm 0.022$ & lattice QCD & \cite{Alexandrou:2021wzv}\\
            $g^s_A$ & $-0.15 \pm 0.09$ & Mominal & \\
        \end{tabular}
    \end{ruledtabular}
\end{table}

\begin{figure}[!h]
    \centering
    \begin{minipage}{0.79\textwidth}
    \centering
        \vspace{0.4cm}
        {$\nu$ differential QE at $E_{\nu}=1\rm GeV $ }
    \includegraphics[width=0.7\textwidth]{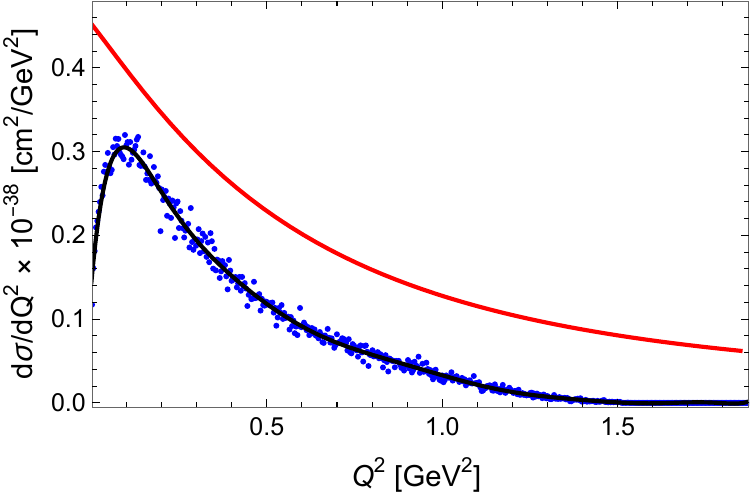}
    \end{minipage}
    \vspace{.5em}
    \caption{ Differential cross sections per nucleon for the neutrino QE NC scattering off Argon at $E_\nu=1$ GeV. 
    We have taken the SM limit with all NSI parameters set to zero.
    The blue dots show the GiBUU prediction for the differential cross section averaged over the nucleons of the Argon, taking the number of ensembles equal to 100. The input form factor parameters are the central values in Tab \ref{tab:QE}. While the black curve is a fit to the blue dots, the red curve shows the prediction for the free nucleons averaged over 18 protons and 22 neutrons
    ({\it i.e.,} the Argon composition) using Eq.~(\ref{eq:diff-sig}).}
    \label{Fig-method}
\end{figure}

The precision of the computation by the GiBUU event generator depends on the input value for the ``number of ensembles". The higher this number, the better is the precision but of course, increasing the number of the ensembles, the time required for running the code also increases. We take the number of the ensembles equal to 100. An example is shown in Fig \ref{Fig-method} for neutrino beam of 1 GeV. The vertical axis is the partial cross section of the neutrino scattering on the nucleons of Argon averaged over protons and neutrons. The blue dots show the results obtained by the GiBUU event generator. The black curve shows the best fit to the blue dots.
The red curve shows the cross section for free nucleons computed using Eq.~(\ref{eq:diff-sig}) and averaged over 18 protons and 22 neutrons composing the Argon nucleus. The significant difference between the GiBUU results and the red curve is of course due to the nuclear effects. As explained in \cite{Leitner:2008ue,Soplin:2023nxg}, the change in the form factors due to the nuclear medium effect is neglected. The three main nuclear medium effects, implemented by GiBUU, are the following:
\begin{itemize}
    \item  Due to the binding energy of the nucleus, the masses of the initial and final nucleons receive a correction of 5\% to 9 \% for nucleon kinetic energy below 1 GeV. This effect alone cannot account for the large difference shown in Fig \ref{Fig-method}.
    \item Fermi blocking of the final nucleon can be significant for momenta below 200~MeV.  The significant suppression for $Q^2<0.1$ GeV$^2$ can be partially attributed to the Fermi blocking but at higher $Q^2$, this effect is negligible. 
    \item The dominant nuclear effect originates from the collision of the ejected nucleon off the rest of nucleons in the nucleus before exiting the nucleus.
    The intuitive reason why this leads to the suppression of cross section is the following. In general, the cross section of $A+B \to C+D$  in the center of mass frame is proportional to $[1/(E_AE_B v_r)][p_C/(E_A+E_B)]$ times the square of the matrix element.  The multiple collisions of the final states off other nucleons resemble carrying them along. We therefore expect an increase in $m_C$ and $E_C$ which will in turn decrease the cross section.
\end{itemize}
The cross section computed by GiBUU can be in general written as 
\begin{align} 
    \left.\frac{d\sigma}{dQ^2}\right|_{\rm GiBUU}
    =\alpha(Q^2,E_\nu, \tilde{F}_i^N ,\tilde{F}_A^N) \left. \frac{d\sigma(Q^2,E_\nu, \tilde{F}_i^N ,\tilde{F}_A^N)}{dQ^2}\right|_{\rm free} 
    \label{alpha}
\end{align}
where $\sigma |_{\rm free}$ is the cross section given in Eq.~(\ref{eq:diff-sig}) without nuclear effects. $\alpha$ accounts for the nuclear effects. In the limit that we neglect the effect of the binding energy on the effective mass of nucleons, $\alpha$ becomes independent of $\tilde{F}_A^N$ and $\tilde{F}_i^N$. Indeed, varying $\tilde{F}_A^N$ of order 1, we found that the variation of $\alpha$ is less than 10 \%. As a result, for small variations of $\tilde{F}_A^N$ and $\tilde{F}_i^N$ (caused by {\it e.g.,} $g_A^s$ or uncertainties in other form factor parameters), we can safely assume that $\alpha $ is independent of $\tilde{F}_A^N$.
Taking $\alpha$ independent of $\tilde{F}_A^N$, we can extract its value as a function of $Q^2$ and $E_\nu$ for the central value of $\tilde{F}_A^N$ at a given $\epsilon^{Au}$ and $\epsilon^{Ad}$ and evaluate the effect of each uncertainty on the cross section using Eqs.~(\ref{eq:diff-sig}, \ref{alpha}). For the purpose of determining the uncertainty in the cross section prediction induced by the form factors, this method has two advantages over using GiBUU to compute the cross section at the lower and upper limits of $\tilde{F}_A^N$: 
(1) Computing $\sigma|_{\rm free}$ is much faster than computing $\sigma|_{\rm GiBUU}$. (2)  Due to the limited number of ensembles, GiBUU itself induces an uncertainty that can be mistaken for the uncertainty induced by the uncertainties of form factors. 
The method invoking Eq (\ref{alpha}) circumvents the spurious uncertainty.

\begin{figure}[h!]
    \centering
    \begin{minipage}{0.49\textwidth}
    \centering
        \vspace{0.4cm}
        {$\nu$ QE cross section}
        \includegraphics[width=1\linewidth]{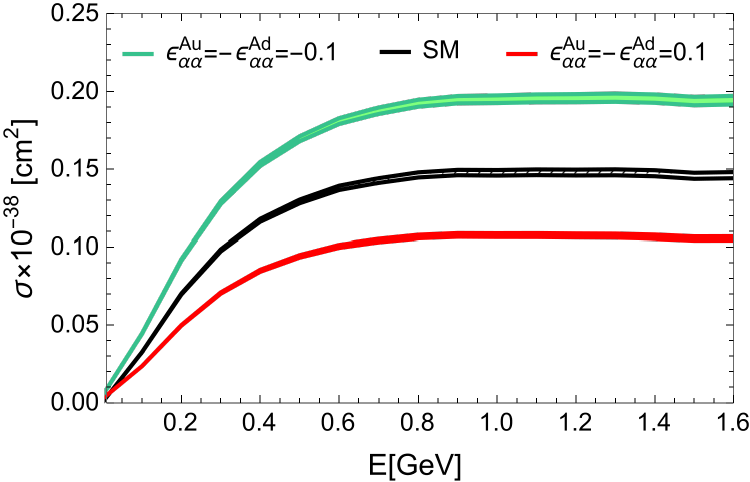}
    \end{minipage}
    \begin{minipage}{0.49\textwidth}
    \centering
        \vspace{0.4cm}
        {$\overline{\nu}$ QE cross section}
        \includegraphics[width=1\linewidth]{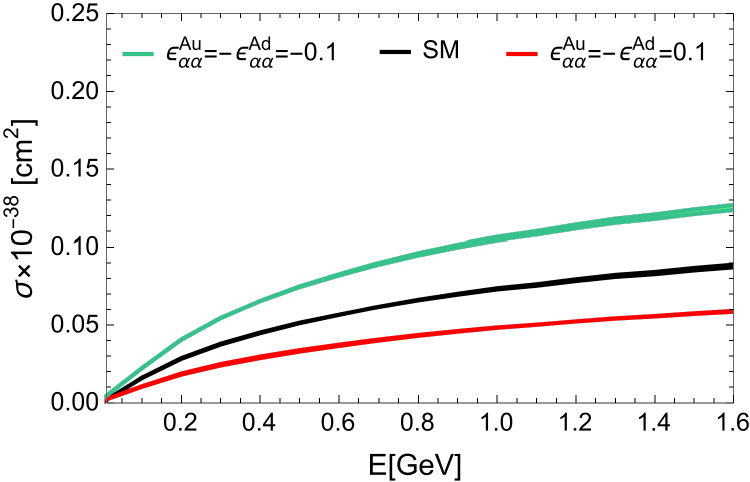}
    \end{minipage}
    \begin{minipage}{0.49\textwidth}
    \centering
        {\vspace{0.4cm}$\nu$  differential QE at $E_\nu=1\ {\rm GeV}$}
        \includegraphics[width=1\linewidth]{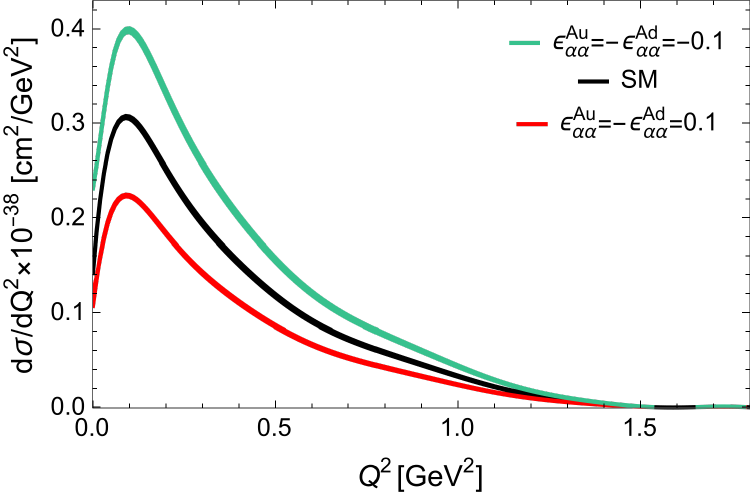}
    \end{minipage}
    \begin{minipage}{0.49\textwidth}
    \centering
        {\vspace{0.4cm}$\overline{\nu}$- differential QE at $E_{\overline{\nu}}=1\ {\rm GeV}$}
        \includegraphics[width=1\linewidth]{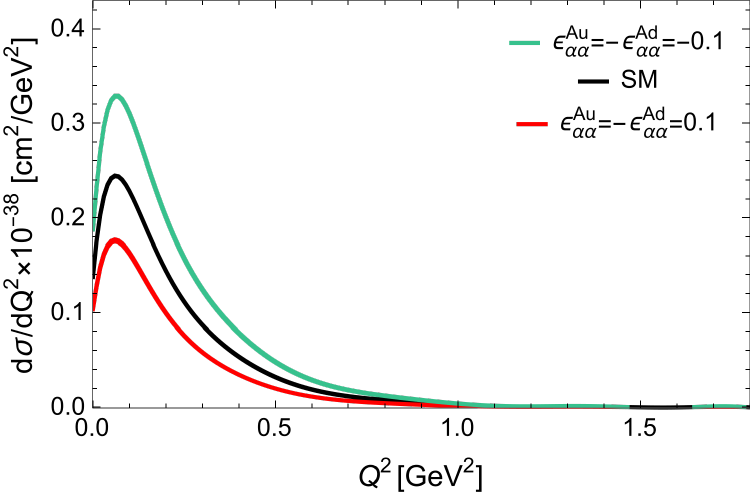}
    \end{minipage}
    \caption{QE scattering cross section for SM and isovector axial NSI. The green and red bands  respectively correspond to $\epsilon_{\alpha\alpha}^{Au}=-\epsilon_{\alpha\alpha}^{Ad}=-0.1$ and  $\epsilon_{\alpha\alpha}^{Au}=-\epsilon_{\alpha\alpha}^{Ad}=0.1$ which saturate the present bounds.  The rest of the NSI parameters are fixed to zero. The black bands show the SM prediction.  The input form factors are shown in Tab.~\ref{tab:QE}. We have varied $m_A$ and  $g_A^s$ within the ranges shown in Tab.~\ref{tab:QE}. The vertical axis shows the cross section per nucleon for scattering off Argon. 
    The left (right) panels are for neutrinos (antineutrinos). The lower panels show the differential cross section at $E_\nu=1$ GeV.}
    \label{fig:DiagIsoViolateQE}
\end{figure}

The impact of vector NSI on the QE NC interaction has already been studied in Ref.~\cite{Papoulias:2016edm}.
Our focus in this section is on the axial NSI. 
We first start with the isospin violating case, which is very constrained.
Fig.~\ref{fig:DiagIsoViolateQE} shows $\stackrel{(-)}{\nu}$-N NC QE scattering cross section with $\epsilon^{Au}_{\alpha \alpha}=-\epsilon^{Ad}_{\alpha \alpha}=\pm 0.1$.
The cross section of scattering off Argon is divided by 40 (its mass number) to give the cross section per nucleon. The left and right panels are respectively for neutrino and antineutrino scatterings. While the lower panels show the differential cross section ($d\sigma/dQ^2$) at $E_\nu=1\ \rm GeV$, the upper panels display the cross section ($\sigma$) versus the incoming neutrino energy, $E_\nu$. The black, green and red bands respectively correspond to the SM, 
$\epsilon^{Au}_{\alpha \alpha}=-\epsilon^{Ad}_{\alpha \alpha}=-0.1$ and $\epsilon^{Au}_{\alpha \alpha}=-\epsilon^{Ad}_{\alpha \alpha}= 0.1$. Neither the SM prediction nor the isovector NSI prediction is sensitive to $g_A^{(8)}$ or $M_A^{(8)}$. The tiny widths of the bands are due to varying $m_A$ and $g_A^s$ as shown in Tab.~\ref{tab:QE}.
It is not surprising that for $Q^2<1$ GeV$^2$, the variation due to $m_A$ is very tiny. As seen from the upper panels, the width of the bands increases at higher energies. This is due to the fact that for larger $E_\nu$, $Q^2$ can have larger values so the factor $Q^2/M_A^2$ in Eq.~(\ref{eq:FA-}) can be significant.
The variation due to the $g_A^s$ uncertainty is surprisingly low for both the SM prediction and the isovector NSI prediction. This is due to the fact that the variations in the contributions of protons and neutrons partially cancel each other.  From Eqs.~(\ref{eq:FAp}, \ref{eq:FAn}), it is manifest that for the SM as well as for the isovector axial NSI, the central values of $\tilde{F}_A^p$ and $\tilde{F}_A^n$
are opposite to each other so the variation of $F_A^s$ (or equivalently of $g^s_A$)
cannot lead to a significant change in the scattering cross section off Argon.
We therefore conclude that when extracting information on the isovector axial NSI from the scattering of neutrinos or antineutrinos off an (almost) isospin singlet nucleus ({\it i.e.,} a nucleus with $Z\simeq A/2$), the uncertainties in the form factors will not be a source of degeneracy even for $-0.1< \epsilon^{Au}=-\epsilon^{Ad}< 0.1$ as long as their magnitudes are larger than about 0.01 which is the limit of reliability of using CC interactions to derive $g_A$ (i.e., the validity of the isospin symmetry). 
Notice that this conclusion also applies to the lepton flavor violating case ($\epsilon^{Au}_{\alpha \beta}=-\epsilon^{Au}_{\alpha \beta}|_{\alpha \ne \beta} $) for which according to Eqs. (\ref{eq:FAp}, \ref{eq:FAn}), the dependence on $F_A^s$ and therefore on $g_A^s$ disappears altogether.
We observe that depending on the sign of $\epsilon^{Au}_{\alpha\alpha}=-\epsilon^{Ad}_{\alpha\alpha}$, we can have a deficit or an excess so the sign of the NSI coupling can also be derived for the isovector axial NSI. This is due to the interference between the SM contribution and the lepton flavor conserving NSI contribution. For the lepton flavor violating case ($\epsilon^{Aq}_{\alpha \beta}$ with $\alpha \ne \beta$), we expect excess in the NC cross section (when summed over the flavor of the final neutrinos) regardless of the sign of the NSI coupling. From Fig.~\ref{fig:DiagIsoViolateQE}, we see that neutrinos and antineutrinos are both sensitive to NSI but the sensitivity of neutrinos is slightly higher.

\begin{figure}[!h]
    \centering
    \begin{minipage}{0.49\textwidth}
    \centering
        \vspace{0.4cm}
        {$\nu$ QE cross section}
        \includegraphics[width=1\linewidth]{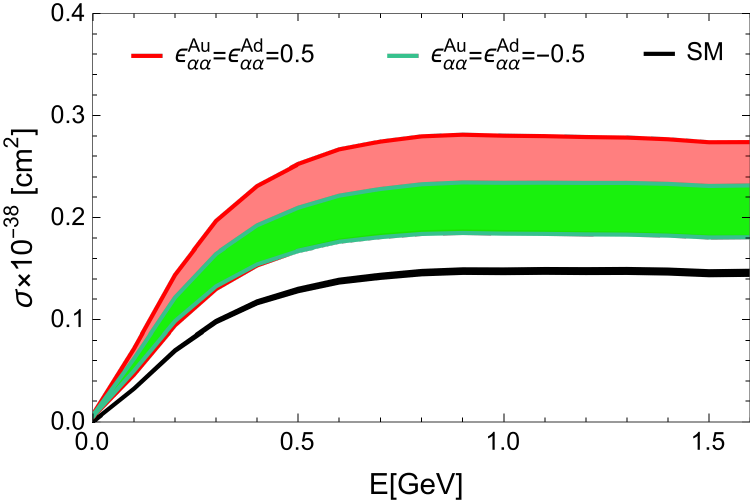}
    \end{minipage}
    \begin{minipage}{0.49\textwidth}
    \centering
        \vspace{0.4cm}
        {$\overline{\nu}$ QE cross section}
        \includegraphics[width=1\linewidth]{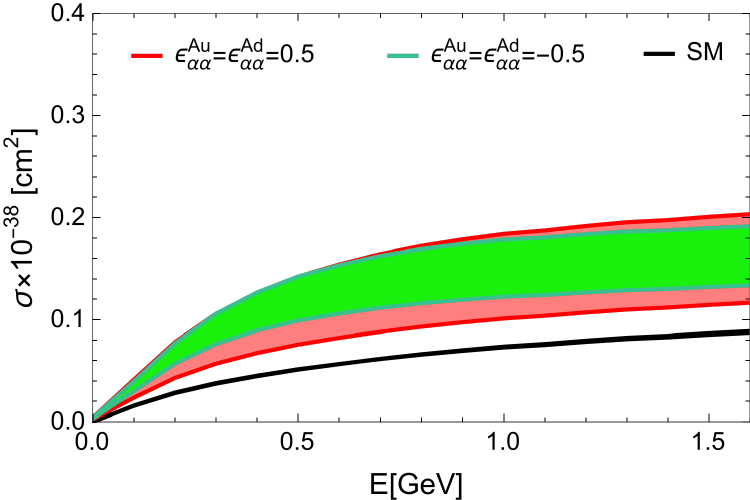}
    \end{minipage}
    \begin{minipage}{0.49\textwidth}
    \centering
        {\vspace{0.4cm}$\nu$ differential QE at $E_\nu=1\ {\rm GeV}$}
        \includegraphics[width=1\linewidth]{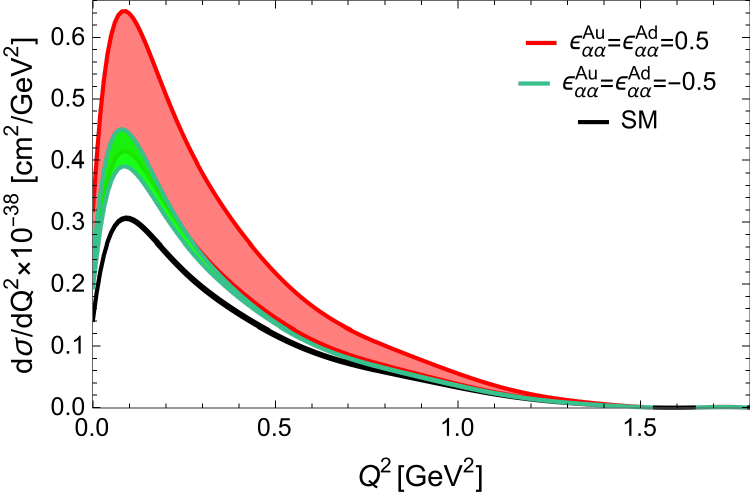}
    \end{minipage}
    \begin{minipage}{0.49\textwidth}
    \centering
        {\vspace{0.4cm}$\overline{\nu}$ differential QE at $E_{\overline{\nu}}=1\ {\rm GeV}$}
        \includegraphics[width=1\linewidth]{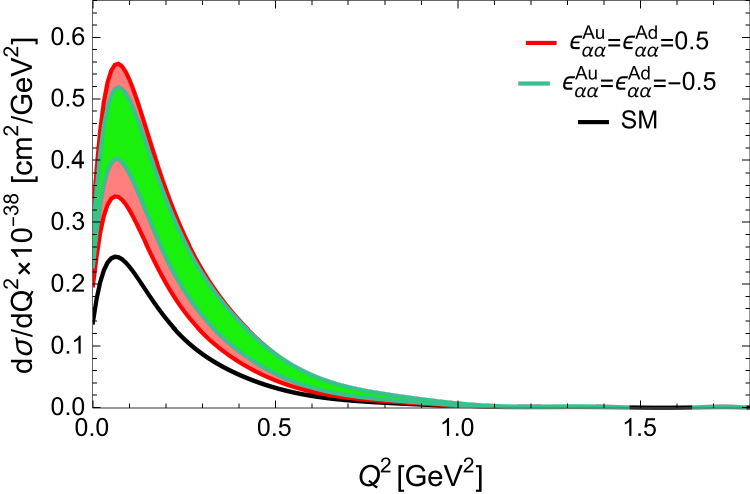}
    \end{minipage}
    \caption{ The same as Fig \ref{fig:DiagIsoViolateQE} except that we have taken  isoscalar NSI with 
 $\epsilon_{\alpha \alpha}^{Au}=\epsilon^{Ad}_{\alpha\alpha}=\pm 0.5$. To obtain the bands, in addition to $m_A$ and $g_A^s$,  the $m^{(8)}_A$ and $g^{(8)}_A$
parameters are also varied within the range shown in Tab.~\ref{tab:QE}.}
    \label{fig:DiagIsoQE}
\end{figure}

\begin{figure}[!h]
    \centering
    \begin{minipage}{0.59\textwidth}
    \centering
        \vspace{0.4cm}
        {$\nu$ differential QE at $E_{\nu}=1\rm GeV $ }
        \includegraphics[width=1\linewidth]{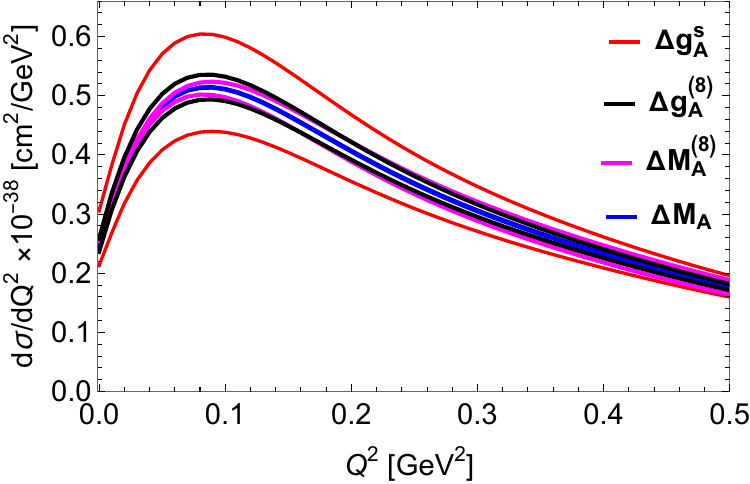}
    \end{minipage}
    \caption{Differential cross sections per nucleon for QE NC scattering off Argon at an energy of $E_\nu = 1$ GeV, taking $\epsilon^{Au}_{\alpha\alpha}=\epsilon^{Ad}_{\alpha\alpha}=0.5$. The blue, magenta, black and red curves respectively show the uncertainty induced by $ M_A$, $ M_A^{(8)}$, $ g_A^{(8)}$ and  $g_A^s$ as they vary within the ranges shown in Tab.~\ref{tab:QE}.}
    \label{fig:DiagIsoMinusQE}
\end{figure}
\begin{figure}[!h]
    \centering
    \begin{minipage}{0.49\textwidth}
    \centering
        \vspace{0.4cm}
        {$\nu$ QE cross section}
        \includegraphics[width=1\linewidth]{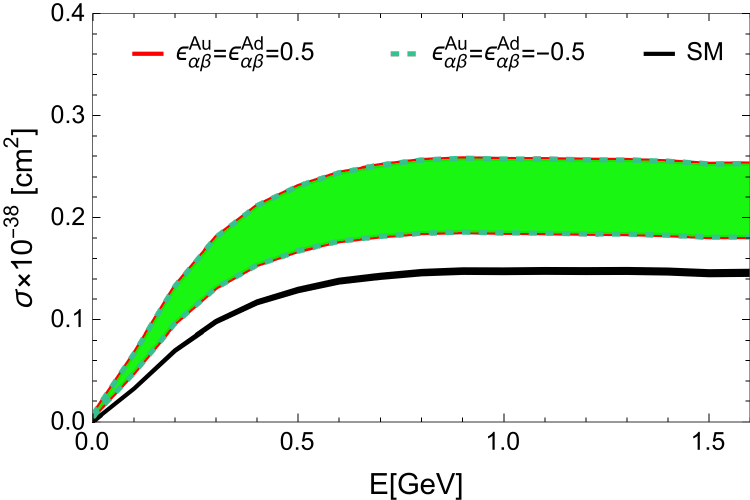}
    \end{minipage}
    \begin{minipage}{0.49\textwidth}
    \centering
        \vspace{0.4cm}
        {$\overline{\nu}$ QE cross section}
        \includegraphics[width=1\linewidth]{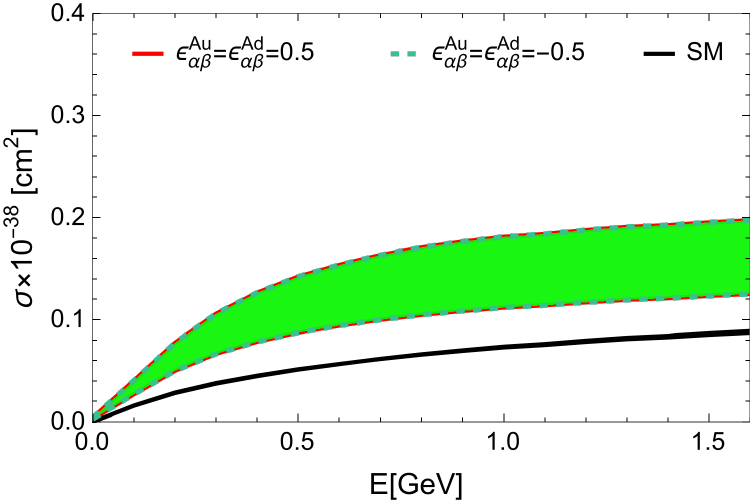}
    \end{minipage}
    \begin{minipage}{0.49\textwidth}
    \centering
        {\vspace{0.4cm}$\nu$ differential QE at $E_\nu=1\ {\rm GeV}$}
        \includegraphics[width=1\linewidth]{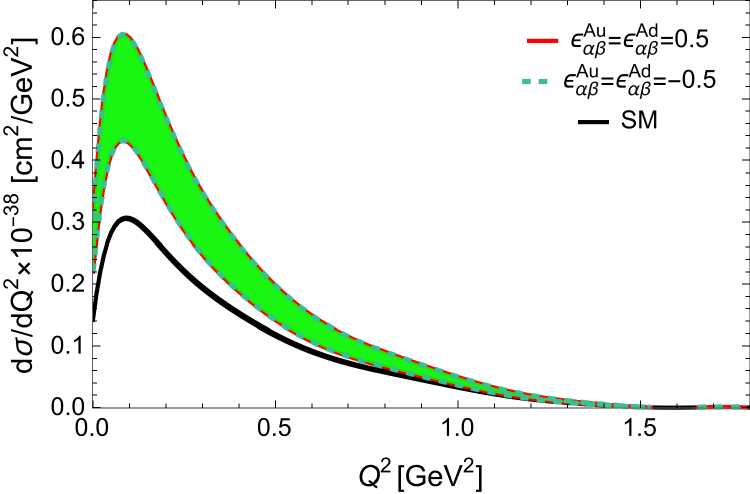}
    \end{minipage}
    \begin{minipage}{0.49\textwidth}
    \centering
        {\vspace{0.4cm}$\overline{\nu}$ differential QE at $E_{\overline{\nu}}=1\ {\rm GeV}$}
        \includegraphics[width=1\linewidth]{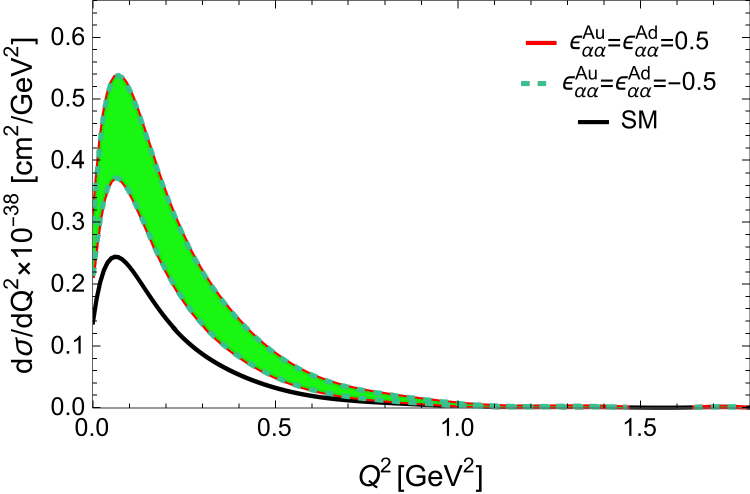}
    \end{minipage}
    \caption{ The same as Fig \ref{fig:DiagIsoViolateQE} except that we have taken lepton flavor violating  isoscalar NSI with 
 $\epsilon_{\alpha \beta}^{Au}=\epsilon^{Ad}_{\alpha\beta}|_{\alpha \ne \beta}=\pm 0.5$. As  expected the  $\epsilon_{\alpha \beta}^{Au}=\epsilon^{Ad}_{\alpha\beta}|_{\alpha \ne \beta}=\pm 0.5$ bands completely overlap. To obtain the bands, in addition to $m_A$ and $g_A^s$,  the $m^{(8)}_A$ and $g^{(8)}_A$
parameters are also varied within the range shown in Tab.~\ref{tab:QE}.}
    \label{fig:OffIsoQE}
\end{figure}
Let us now discuss the isospin invariant case which is less constrained as discussed in Sect.~\ref{sec:constraints}. 
Fig.~\ref{fig:DiagIsoQE} is similar to Fig.~\ref{fig:DiagIsoViolateQE} except that we have set $\epsilon^{Au}_{\alpha \alpha}=\epsilon^{Ad}_{\alpha \alpha}=\pm 0.5$.
The two NSI bands with opposite signs $\epsilon_{\alpha \alpha}^{Au}=
\epsilon_{\alpha \alpha}^{Ad}=\pm 0.5$ overlap with each other which means that the scattering cross section cannot determine the sign of the NSI coupling. Unlike the case of the isovector axial NSI, here the widths of NSI are large which means the form factor parameters induce large uncertainties in the cross section. Remember that the cross section for the isoscalar axial NSI depends on $g_A^{(8)}$ and $m_A^{(8)}$, too. To obtain these bands, we have simultaneously varied all $g_A^{(8)}$, $m_A^{(8)}$, $g_A^{s}$ and $m_A$ within the ranges in Tab~\ref{tab:QE}. To disentangle the uncertainty induced by each form factor parameter, we have included Fig \ref{fig:DiagIsoMinusQE} in which each pair of curves show the uncertainty induced by a single form factor parameter, setting the rest of the parameters equal to their central values. Unlike the case of the isovector NSI, the uncertainty induced by $g_A^s$ is significant. This is because at $\epsilon_{\alpha \alpha}^{Au}=\epsilon_{\alpha \alpha}^{Ad}=\pm 0.5$, the central values of $\tilde{F}_A^p$ and $\tilde{F}_A^n$ are not opposite to each other and cancellations between the variations of the proton and neutron contribution do not take place.

If the excess relative to the SM is larger than 30\%, comparing Figs.~(\ref{fig:DiagIsoViolateQE}, \ref{fig:DiagIsoQE}),  we can conclude that it is induced by isoscalar axial NSI because the bounds on the isovector NSI cannot allow such a large deviation. However, because of the large uncertainties in the cross section prediction induced by $g_A^s$ for the case of isoscalar NSI, the derivation of the value of $\epsilon^{Au}+\epsilon^{Ad}$ will suffer from large uncertainty. If such an excess is observed it would be imperative to improve the determination of $g_A^s$  which is fortunately possible thanks to the lattice QCD method \cite{Alexandrou:2019brg}.
For isoscalar NSI, we only expect excess. For the isovector NSI with $0<\epsilon^{Au}_{\alpha \alpha }=-\epsilon^{Ad}_{\alpha \alpha }<0.1$, we expect a deficit up to 30 \%. If the deficit turns out to be larger, we should find a beyond SM solution other than NSI because the bounds on $\epsilon^{Au}-\epsilon^{Ad}$ cannot allow a deficit larger than 30 \%. However, a deficit of $\sim 30$\% or less can be considered as a hint for $0<\epsilon^{Au}_{\alpha \alpha }=-\epsilon^{Ad}_{\alpha \alpha }<0.1$.

Fig.~\ref{fig:OffIsoQE} shows how the QE scattering cross section changes in the presence of lepton flavor violating NSI. Notice that in this case, the green/red curves show the sum of the cross sections of $\nu_\alpha +N \to \nu_\alpha +N $ (from the SM) and of
$\nu_\alpha +N \to \nu_\beta +N $ (from the lepton flavor violating NSI). The latter is proportional to $
|\epsilon_{\alpha \beta}^{Au}|^2$ and is therefore always positive and independent of the sign of $\epsilon_{\alpha \beta}^{Au}=\epsilon_{\alpha \beta}^{Ad}$.  As a result, the red and green curves have complete overlap. Notice that  the cross section of $\nu_\alpha +N \to \nu_\beta +N$ does not depend on $g_A$   so the dependence of $\sigma(\nu_\alpha +N \to \nu_\beta +N$) on $m_A$ is only through the $Q^2$ dependence of $F_A^s$. 
 The dominant uncertainty source is $g_A^{(8)}$ and $M_A^{(8)}$. Despite the large uncertainty in the cross section of $\nu_\alpha +N \to \nu_\beta +N$, the deviation from SM is distinguishable. 

\begin{figure}[!h]
   \centering
   \begin{minipage}{0.49\textwidth}
   \centering
        \vspace{0.4cm}
        {$\nu$ QE cross section}
       \includegraphics[width=1\linewidth]{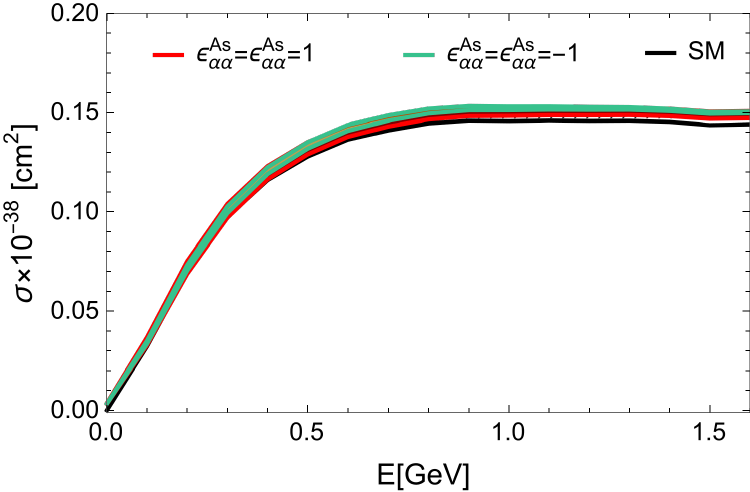}
   \end{minipage}
   \begin{minipage}{0.49\textwidth}
    \centering
        \vspace{0.4cm}
        {$\overline{\nu}$ QE cross section}
       \includegraphics[width=1\linewidth]{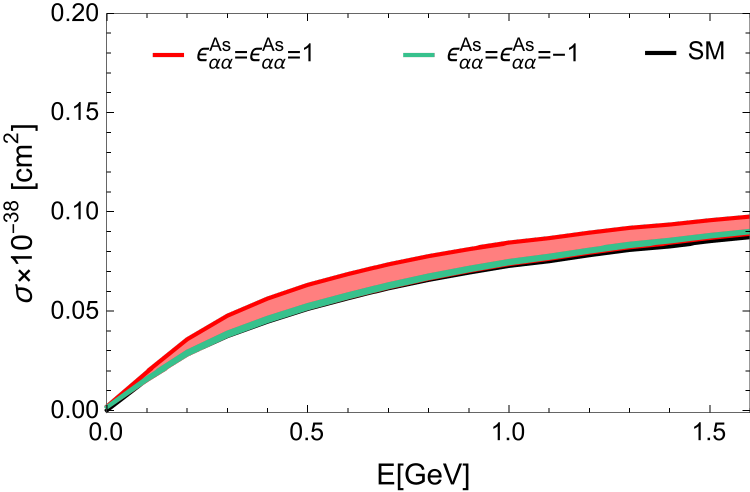}
    \end{minipage}
   \begin{minipage}{0.49\textwidth}
    \centering
        {\vspace{0.4cm}$\nu$ differential QE at $E_\nu=1\ {\rm GeV}$}
        \includegraphics[width=1\linewidth]{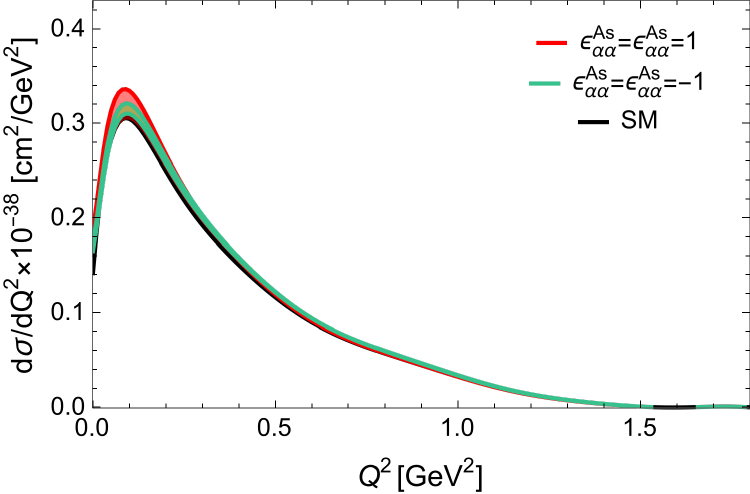}
    \end{minipage}
    \begin{minipage}{0.49\textwidth}
    \centering
        {\vspace{0.4cm}$\overline{\nu}$  differential  QE at $E_{\overline{\nu}}=1\ {\rm GeV}$}
        \includegraphics[width=1\linewidth]{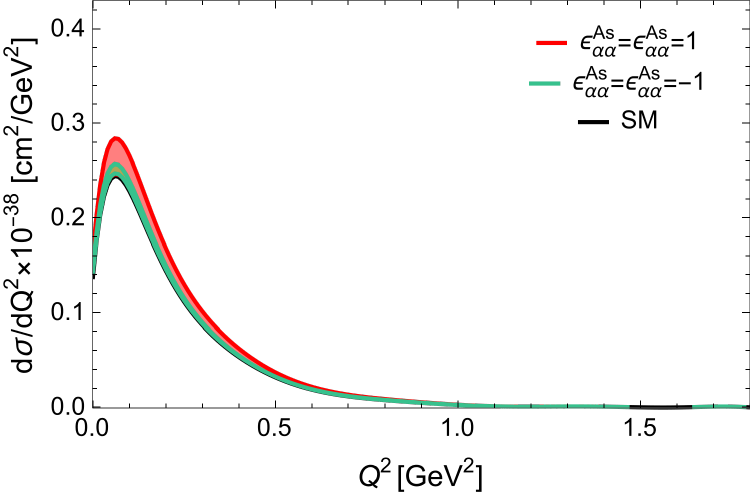}
   \end{minipage}
    \caption{ The same as Fig \ref{fig:DiagIsoViolateQE} except that we have taken  isoscalar NSI $\epsilon_{\alpha \alpha}^{As}=\pm 1$. }
    \label{fig:StrangeQE}
\end{figure}

Fig.~\ref{fig:StrangeQE} shows the impact of $\epsilon^{As}_{\alpha\alpha}=\pm 1$ on the cross section. As seen from the figure, the sensitivity to $\epsilon^{As}_{\alpha\alpha}$ is too small to be resolvable.
There are two reasons for the suppressed sensitivity: (1) The effect is suppressed by the coefficient $F_A^s\simeq -0.15$; (2) The variations due to $\epsilon^{As}$ in the cross sections of neutron and proton are in opposite directions so they cancel each other. Scattering off free protons will therefore be more sensitive to $\epsilon^{As}$ than scattering off an (almost) isosinglet nucleus.

 \begin{table}[!h]
    \caption{The form factor parameters for $\stackrel{(-)}{\nu}-N$ NC resonance scattering\label{tab:Res}}
   \begin{ruledtabular}
\begin{tabular}{ccccccc} 
            Parameter & $m_V^{\Delta}$ & $m_A^{\Delta}$ & $C_{3}^{V}(0)$ & $C_{4}^{V}(0)$ & $C_{5}^{V}(0)$ &  $C_{5}^{A}(0)$\\
            \hline
             Value & $0.84\ \rm GeV$ & $0.94 \pm 0.03\ \rm GeV$ & $2.13$ & $-1.51$ &  $0.48$ & $1.19 \pm 0.08$ \\
             \hline
             References & \cite{Paschos:2003qr,Lalakulich:2006sw} & \cite{Graczyk:2009qm} & \cite{Lalakulich:2006sw} & \cite{Lalakulich:2006sw} & \cite{Lalakulich:2006sw} & \cite{Graczyk:2009qm}\\            
        \end{tabular}
    \end{ruledtabular}
\end{table}

\begin{figure}[h!]
    \centering
    \begin{minipage}{0.49\textwidth}
    \centering
        \vspace{0.4cm}
        {\centering{$\nu$ resonance cross section}}
        \includegraphics[width=1\linewidth]{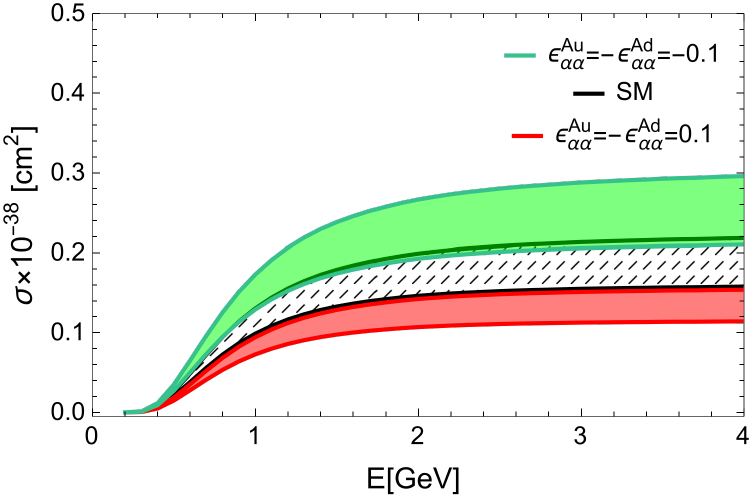}
    \end{minipage}
    \begin{minipage}{0.49\textwidth}
    \centering
        \vspace{0.4cm}
        {$\overline{\nu}$ resonance cross section}
        \includegraphics[width=1\linewidth]{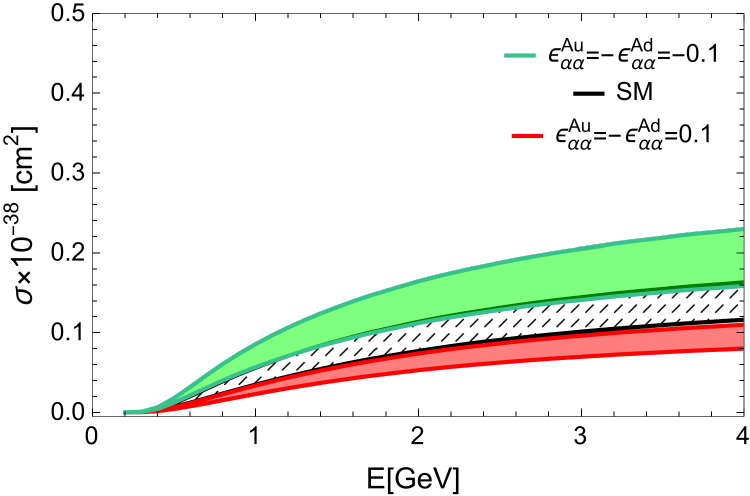}
    \end{minipage}
    \begin{minipage}{0.49\textwidth}
    \centering
        \vspace{0.45cm} $\nu$  differential resonance at $E_\nu=2\ {\rm GeV}$
        \includegraphics[width=1\linewidth]{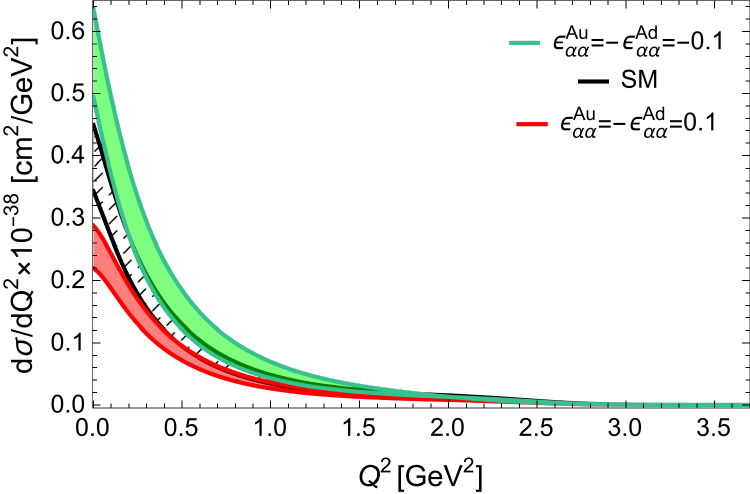}
    \end{minipage}
    \begin{minipage}{0.49\textwidth}
    \centering
        \vspace{0.4cm}$\overline{\nu}$- differential resonance at $E_{\overline{\nu}}=2\ {\rm GeV}$
        \includegraphics[width=1\linewidth]{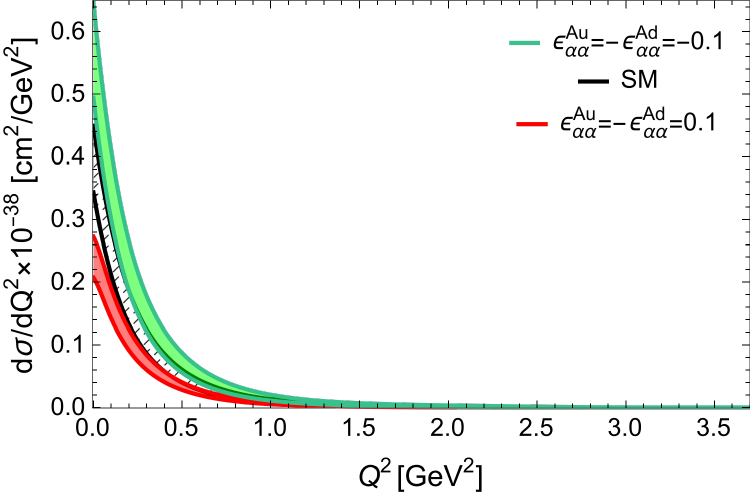}
    \end{minipage}
    \caption{Resonance scattering cross section for SM and isovector axial NSI. The green and red bands  respectively correspond to $\epsilon_{\alpha\alpha}^{Au}=-\epsilon_{\alpha\alpha}^{Ad}=-0.1$ and  $\epsilon_{\alpha\alpha}^{Au}=-\epsilon_{\alpha\alpha}^{Ad}=0.1$ which saturate the present bounds.  The rest of the NSI parameters are fixed to zero. The black bands show the SM prediction.  The input form factors are shown in Tab.~\ref{tab:Res}. We have varied $m_A^{\Delta}$ and  $C_{5}^{A}(0)$ within the ranges shown in Tab.~\ref{tab:Res}. The vertical axis shows the cross section per nucleon for scattering off Argon. 
    The left (right) panels are for neutrinos (antineutrinos). The lower panels show the differential cross section at $E_\nu=2$ GeV.}
    \label{fig:DiagIsoViolateRes}
\end{figure}

Let us now discuss the resonance scattering which receives contributions only from the isovector NSI (i.e., $\epsilon^{Au}=- \epsilon^{Ad} \neq 0$). Table \ref{tab:Res} shows the relevant form factor parameters which we take as the input for GiBUU. The detailed definitions of the parameters and relevant relations are given in \cite{Lalakulich:2006sw}. Refs. \cite{Paschos:2003qr,Lalakulich:2006sw} do not report any error bars for the vector parameters. However, Ref. \cite{Graczyk:2009qm}, invoking the input for $C_V$ and $m_V^\Delta$ from \cite{Lalakulich:2006sw}, derives $C_5^A (0)$ and $m_A^\Delta$ with uncertainties shown in table \ref{tab:Res}. Notice that the observables used to derive the parameters in table \ref{tab:Res} (i.e., electromagnetic interaction of charged lepton and CC scattering of neutrinos) are both free from any contamination due to NSI. From Eq. (\ref{CAiCORRECTED}), we observe that only for $\epsilon^{Au}- \epsilon^{Ad} > 0.1$, the NSI effects can be distinguished from the uncertainties in $C_5^A (0)$. Fig. \ref{fig:DiagIsoViolateRes} demonstrates this expectation. We use GiBUU to compute the cross section for the central values of $C_5^A$ and $m_A^\Delta$. Similarly to the case of QE scattering, we take the ratio of the cross section in nuclear media to that of free nucleons independent of the form factors. Indeed, such an assumption is explicitly used in extracting the form factors from the Deuteron in the literature \cite{Graczyk:2009qm}. As seen from the figure, for $\epsilon^{Au}=- \epsilon^{Ad} =0.1$, 
the NSI bands barely touch the SM band which means in the optimistic case that the uncertainties are only due to  $C^A_5$ and $m_A^\Delta$, there is hope to improve the bounds on $\epsilon^{Au}=- \epsilon^{Ad}$ from the resonant scattering. We examined the possibility of including an uncertainty of 5 \% in $m_V^\Delta$ and $C_i^V$ ({\it i.e.,} uncertainties achievable by lattice QCD method; see {\it e.g.,} Tab.~\ref{tab:QE}) and found that the widening of the bands is small. This means our conclusion remains valid in the presence of $C^A_5$ and $m_A^\Delta$  uncertainties at the level of the uncertainties achievable by the lattice QCD.

\section{Summary}
\label{summary}
We have started with reviewing the present bounds on the NSI couplings from various experiments with special attention to the possible degeneracies. 
We have pointed out that there is an approximate degeneracy between the SM solution with $\epsilon^{{A/V} q}_{\alpha \beta}=0$ and a non-trivial solution with $\epsilon_{\alpha \beta}^{{A/V} q}=\epsilon_{ee}^{{A/V} q} \delta_{\alpha \beta}$ and with values shown in Eq.~(\ref{eq:degenerate}). If we add  $\epsilon^{{A/V}s}_{\alpha \beta}=\epsilon^{{A/V}s}_{ee}\delta_{\alpha \beta}$ to the previous relations, the hadronic current $J^\mu_{\rm had}$ (see Eqs.~(\ref{eq:Jahed}-\ref{eq:ASM}) for the definition of $J^\mu_{\rm had}$) transforms to $-J^\mu_{\rm had}$.
As a result,  neither the scattering experiments nor the neutrino oscillation experiments can distinguish this non-trivial solution from the SM. We then focused on the values of NSI coupling around the SM solution and enumerated the bounds on the couplings presented in the literature. 
Despite stringent bounds on vector NSI ($\epsilon_{\alpha \beta}^{Vq}<0.05$, $q=u,d$), the axial NSI couplings $\epsilon_{\tau\tau}^{Au}+\epsilon_{\tau\tau}^{Ad}$, $\epsilon_{e\tau}^{Au}+\epsilon_{e\tau}^{Ad}$ and $\epsilon_{ee}^{Au}+\epsilon_{ee}^{Ad}$ as well as $\epsilon^{As}_{\alpha \beta}$ can be as large as $\mathcal{O}(1)$. Moreover,  $\epsilon_{\tau\tau}^{Au}-\epsilon_{\tau\tau}^{Ad}$ and $\epsilon_{ee}^{Au}-\epsilon_{ee}^{Ad}$ 
can be as large as 0.2.

We have then discussed degeneracies in extracting the values of the form factors and NSI couplings from QE and resonance neutrino scattering off nuclei. We have discussed how these degeneracies can be solved by employing extra information on the value of the form factors from the lattice QCD prediction or from the scattering of charged leptons off nuclei. In particular, we have highlighted two new avenues to derive information on NSI: (1) While the lattice QCD computation shows $ F_1^s(Q^2)<0.01$, nonzero $\epsilon^{Vu}$ or $\epsilon^{Vd}$ can fake  $F_1^s\sim 0.01$. The $\nu_\mu$ scattering in MiniBooNE already constraints $|F_1^s(Q^2)|$ for 
$Q^2<1$~GeV$^2$ to be smaller than 0.1. If future neutrino scattering experiments, $\nu_\alpha +N \to \nu_\alpha+N$ yield $F_1^s(Q^2)$ to be $\mathcal{O}(0.01)$, it should be interpreted as a nonzero $2\epsilon_{\alpha \alpha}^{Vu}+\epsilon_{\alpha \alpha}^{Vd}\sim 0.01$, close to the bound from neutrino oscillation and CE$\nu$NS data \cite{Coloma:2023ixt}. (2) Low energy (50~MeV$<E_\nu<$1~GeV) atmospheric neutrino data collected by the KamLAND experiment, which is used to constrain $F_A^s$, can already set a bound on $|\epsilon_{\tau\tau}^{Au}+\epsilon_{\tau\tau}^{Ad}|$ stronger than the present limits. More accurate measurement of low energy atmospheric neutrino data by future detectors such as JUNO can yield valuable information on $\epsilon_{\tau\tau}^{Au}+\epsilon_{\tau\tau}^{Ad}$ and/or $\epsilon_{\tau\tau(e)}^{Au}+\epsilon_{\tau\tau(e)}^{Ad}$.

We have used the GiBUU event generator to compute  the 
differential and total
cross sections of QE and resonance NC interactions of neutrinos and antineutrinos off nucleons in Argon, taking into account the nuclear effects. 
 We have found that while an isoscalar axial NSI, $\epsilon^{Au}=\epsilon^{Ad}$ leads to only an excess in the QE NC cross section relative to the SM prediction,  the lepton flavor conserving isovector axial NSI, $\epsilon^{Au}_{\alpha \alpha}=-\epsilon^{Aa}_{\alpha \alpha}
$ can lead to both excess or deficit depending on its sign. Considering the bounds on $\epsilon^{Au}-\epsilon^{Ad}$, the deviation for the isovector case cannot be larger than 30 \%. Thus, an excess of more than 30 \% can be interpreted as axial isoscalar NSI (but not isovector NSI). A deficit of more than 30 \% should find another beyond standard model explanation.   We have found that in the presence of isoscalar axial NSI, the uncertainties of the form factors (mainly that of $F_A^s$) induce large uncertainty in the prediction of the QE NC cross section.  However, for the isovector NSI, the uncertainties induced by form factors on the cross sections of (anti)neutrinos off neutron and proton cancel each other. Observing a deficit of $<30 \%$ in the cross section will allow us 
to determine a positive value of $\epsilon^{Au}_{\alpha\alpha}=-\epsilon^{Ad}_{\alpha\alpha}$ free from the form factor uncertainties.
With an excess less than $\sim 30\%$ in the cross section of the QE (anti)neutrino scattering, we cannot determine if $\epsilon^{Au}=\epsilon^{Ad}$ or not. Considering that the isoscalar NSI cannot affect the $\Delta$ resonance cross section, the measurement of its cross section can be used as a discriminant between isovector and isoscalar NSI.

Finally, we have studied the impact of $\epsilon^{As}$ on the cross sections. Even with 
$\mathcal{O}(1)$ values, its effects on the cross section of neutrinos or antineutrinos off Argon will be buried in the form factor uncertainties because (1) its effects are suppressed by $F_A^s$; (2) the effects on neutron and protons cancel each other. Because of the second reason, it seems the scattering off free proton has a better chance of determining $\epsilon^{As}$ than scattering off heavy nuclei with $A\simeq 2Z$. 

\appendix

\section{Form factors for standard NC QE scattering}\label{sect:App-a}

To compute the amplitude of the QE scattering of neutrinos, we need to sandwich the hadron current appearing in Eq.~(\ref{eq:JJ}) between the initial and final nucleon states. Decomposing the hadronic current of the SM as $V_{SM}^\mu+A_{SM}^\mu$ (see Eqs.~(\ref{eq:Jahed}, \ref{eq:VSM}, \ref{eq:ASM})), we can write
\begin{align}
    \langle N(p') | V_{\rm SM}^\mu | N(p) \rangle&=\overline{u}_N(p') \left[\left( \gamma^\mu - \frac{\not{q} q^\mu}{q^2} \right) \tilde{F}_1^N(Q^2) + \frac{i}{2 M_N} \sigma^{\mu \nu} q_\nu \tilde{F}_2^N (Q^2)\right] u_N(p) \label{eq:FFV}
    \\
    \langle N(p') | A_{\rm SM}^\mu | N(p) \rangle &= \overline{u}_N(p') \left[ \gamma^\mu \gamma_5 \tilde{F}_A^N (Q^2)+ \frac{q^\mu}{M_N} \gamma_5 \tilde{F}_P^N(Q^2)]\right] u_N(p) 
    \label{eq:FFA}
\end{align}
with \( q_\mu = p'_\mu- p_\mu\) and \( N = p, n \). Here $ M_N $  and $u_N$ denote the nucleon mass and the spinor describing $N$.  $ \tilde{F}_1^N (Q^2)$ and $ \tilde{F}_2^N (Q^2)$ are the vector form factors. $ \tilde{F}_A^N (Q^2)$  and $ \tilde{F}_P^N (Q^2)$ are respectively known as the axial and pseudoscalar form factors. The operators in $V_{\rm SM}^\mu$ ({\it i.e.,} $\overline{Q}\gamma^\mu \tau^3 Q$, $\overline{Q}\gamma^\mu Q$ and $\overline{s}\gamma^\mu s$ appearing in Eq.~(\ref{eq:VSM})) are the same operators that the electromagnetic current is composed of:
\begin{align}
    J_{\rm EM}^\mu=\frac{2}{3}\overline{u}\gamma^\mu u-\frac{1}{3}\overline{u}\gamma^\mu d-\frac{1}{3}\overline{s}\gamma^\mu s=
    \frac{1}{6}\overline{Q}\gamma^\mu Q+\frac{1}{2}\overline{Q}\gamma^\mu\tau^3 Q-\frac{1}{3}\overline{s}\gamma^\mu s. \label{JEM}
\end{align}
The matrix elements, $\langle N(p')| J_{\rm EM}^\mu |N(p)\rangle$, can  also   be parametrized  as in  Eq.~(\ref{eq:FFV}), replacing $\tilde{F}_i^N(Q^2)\to {F}_i^N(Q^2)$ where $F_i^N$ are the famous Dirac and Pauli form factors. The values of $F_i^N$ are well measured from the scattering of charged leptons off nuclei. It is therefore convenient to write $\tilde{F}^N_i$ in terms of ${F}^N_i$. 
To do so, we will also need the form factors associated to $\langle N(p')| \overline{s} \gamma^\mu s |N(p)\rangle$ which can again be parameterized as in Eq.~(\ref{eq:FFV}) with $\tilde{F}_i^N \to F_i^s$.  Notice that we have implicitly used the isospin symmetry:
\begin{align*}
    \langle p(p')|\overline{s}\gamma^\mu s |p(p) \rangle=\langle n(p')|\overline{s}\gamma^\mu s |n(p) \rangle.
\end{align*}
Moreover, the isospin symmetry implies 
\begin{align*}
    &\langle p(p')|\overline{Q}\gamma^\mu Q |p(p) \rangle=\langle n(p')|\overline{Q}\gamma^\mu Q |n(p) \rangle \\
    \quad {\rm and} \quad &
    \langle p(p')|\overline{Q}\gamma^\mu \tau^3 Q |p(p) \rangle=-\langle n(p')|\overline{Q}\gamma^\mu \tau^3 Q |n(p) \rangle.
\end{align*}

Using these relations and Eq.~(\ref{eq:VSM}), we obtain
\begin{align}
    \tilde{F}_i^p (Q^2)&= \left( \frac{1}{2} - 2 \sin^2 \theta_W \right) F_i^p (Q^2)- \frac{1}{2} F_i^n(Q^2) - \frac{1}{2} F_i^s(Q^2),\quad i=1,2\\
    \tilde{F}_i^n (Q^2)&= \left( \frac{1}{2} - 2 \sin^2 \theta_W \right) F_i^n (Q^2)- \frac{1}{2} F_i^p(Q^2) - \frac{1}{2} F_i^s(Q^2),\quad i=1,2
\end{align}

The Dirac and Pauli form factors, $F_i^N$ are related to the famous electric ($G_E^N$) and magnetic $(G_M^N$) as
\begin{align}
    F_1^N(Q^2)&=\frac{1}{1+\frac{Q^2}{4M_N^2}}\left[\Big(\frac{Q^2}{4M_N^2}\Big) G_M^N(Q^2)+G_E^{N}(Q^2)\right],\\ 
    F_2^N(Q^2)&=\frac{1}{1+\frac{Q^2}{4M_N^2}}\left[G_M^N(Q^2)-G_E^N(Q^2)\right], \\
    F_1^s(Q^2)&=\frac{F_1^s(0)Q^2}{(1+\frac{Q^2}{4M_N^2})(1+\frac{Q^2}{M^2_V})},\\
    F_2^s(Q^2)&=\frac{F_2^s(0)}{(1+\frac{Q^2}{4M_N^2})(1+\frac{Q^2}{M^2_V})}. 
\end{align}
As is well-known, $G_E^N(0)$ is the electric charge of nucleon ($G_E^p(0)=1$ and   $G_E^n(0)=0$). $G_M^N(0)$ is its magnetic moment in terms of the nuclear magneton, $e/(2M_N)$ ($G_M^p(0)=2.79$ and   $G_M^n(0)=-1.91$ \cite{ParticleDataGroup:2024cfk}). These quantities are measured with a mind-blowing accuracy \cite{ParticleDataGroup:2024cfk}. 
However, the determination of the $Q^2$ dependence of the form factor is subject to larger uncertainties. Among the various $Q^2$ parametrizations of these Sachs form factors, we use the BBBA2007 form factors provided in Ref.~\cite{Bodek:2007ym}, which is also the default parametrization in the GiBUU project.

From Eq.~(\ref{eq:ASM}), we observe that while the contributions from the $u$ and $d$ quarks to the proton and the neutron are opposite, the contributions from the strangeness are the same. Thus, \cite{Leitner:2008ue} 
\begin{align}
    \tilde{F}_{A}^{p,n}(Q^2) = \mp \frac{1}{2} F_A (Q^2)+ \frac{1}{2} F_A^s(Q^2),
\end{align}
where the minus sign corresponds to protons and the plus sign corresponds to neutrons. $F_A$ and $F_A^s$ are respectively the contributions from $Q=(u \ d)$ and $s$.

\section{Isospin $3/2$ resonance form factors}\label{app-b}
In this appendix, we review the form factors required to compute the cross section of $\nu+N\to \nu+\Delta$.
For this, we need the matrix elements of the hadronic current \cite{Leitner:2006sp}:
\begin{align}\label{currentd}
\langle \Delta^{+}|J_{\rm had}^{\mu} (0)|p\rangle
= \langle\Delta^{0}|J_{\rm had}^{\mu} (0)|n\rangle
= \overline{\psi}_{\delta}(p^{\prime}) \Gamma^{\delta \mu }_{3/2+}u_N(p).
\end{align}
Here ${\psi}_{\delta}(p^{\prime})$ is the Rarita-Schwinger spinor for the $\Delta$, $\Gamma^{\delta \mu }_{3/2+}$ is the  vertex and  $u_N(p)$ is the Dirac spinor for the nucleon \cite{Leitner:2006sp,Leitner:2008ue,Lalakulich:2005cs}.
The transition vertex  $\Gamma^{\delta \mu }_{3/2+}$ is described in terms of the vector $\mathcal{V}^{\delta \mu}_{3/2}$ and axial transition vertices $\mathcal{A}^{\delta \mu}_{3/2}$ \cite{Leitner:2008ue,Lalakulich:2005cs}\footnote{For resonance with negative parity state $\Gamma^{\delta \mu }_{3/2-}=\Gamma^{\delta \mu }_{3/2+} \gamma^5$.} 
\begin{align}
    \Gamma^{\delta \mu }_{3/2+} = \left[ \mathcal{V}^{\delta \mu }_{3/2} +\mathcal{A}^{\delta \mu }_{3/2}\right] \gamma^{5}.
    \label{eq:AP2-eq7}
\end{align}

The vector and axial components, $\mathcal{V}^{\delta \mu}_{3/2}$ and $\mathcal{A}^{\delta \mu}_{3/2}$,  are described in terms of the vector and axial vector form factors as follows \cite{Lalakulich:2005cs, Leitner:2008ue}:

\begin{equation}
\begin{split}
    \mathcal{V}^{\delta \mu }_{3/2} &=\frac{\tilde{C}_{3}^V(Q^2)}{M_N} (g^{\delta \mu} \slashed{q} - q^{\delta} \gamma^{\mu})+\frac{\tilde{C}_4^V(Q^2)}{M_N^2} (g^{\delta \mu} q\cdot p^{\prime} - q^{\delta} {p^{\prime}}^{\mu})  +  \frac{\tilde{C}_5^V(Q^2)}{M_N^2} (g^{\delta \mu} q\cdot p - q^{\delta} p^{\mu})\\&  + g^{\delta \mu} \tilde{C}_6^V(Q^2),
    \end{split}
\end{equation}
\begin{align}
    \mathcal{A}^{\delta \mu }_{3/2} = \left[\frac{C_3^A(Q^2)}{M_N} (g^{\delta\mu} \slashed{q} - q^{\delta} \gamma^{\mu})+\frac{\tilde{C}_4^A(Q^2)}{M_N^2} (g^{\delta \mu} q\cdot p^{\prime} - q^{\delta} {{p^{\prime}}^{\mu}}) +{\tilde{C}_5^A(Q^2)} g^{\delta \mu}  + \frac{\tilde{C}_6^A(Q^2)}{M_N^2} q^{\delta} q^{\mu}\right] \gamma^{5}, 
\end{align}
where $M_N$ is the nucleon mass. 
In the $\nu+N\to \nu+\Delta$ process, the isospin of the baryon is changed by one unit so it cannot receive any contribution from the isospin invariant operators $\overline{Q} \gamma^\mu Q$, $\overline{Q} \gamma^\mu \gamma^5 Q$, $\overline{s}\gamma^\mu s$ and $\overline{s}\gamma^\mu \gamma^5 s$. This transition is only sensitive to the isovector part of $J_{\rm had}^\mu$  which is shown below \cite{Leitner:2006sp,Leitner:2008ue}
\begin{align}\label{AP2-eq3}
    \left( 1 - 2 \sin^2 \theta_W \right) \frac{1}{2}\overline{Q}\gamma^\mu\tau^3 Q- \frac{1}{2}\overline{Q}\gamma^\mu\gamma^5 \tau^3 Q .
\end{align}
Similarly, the amplitude of $e+N \to e+\Delta$ or of $\mu+N \to \mu+\Delta$ receives only a contribution from the isovector part of the electromagnetic current shown in Eq.~(\ref{JEM}): $\overline{Q}\gamma^\mu \tau^3 Q$.
It is therefore straightforward to relate $\tilde{C}_i^V$ to $C_i^N$ which are obtained from the resonant interaction of the charged leptons with nucleons \cite{Leitner:2006sp,Leitner:2008ue}:
\begin{align}
    \tilde{C}_i^V(Q^2) = -(1 - 2 \sin^2 \theta_W) C_i^N(Q^2),\qquad N=p,n.
\end{align}
The isospin symmetry relates the matrix elements of axial isovector operator $\overline{Q} \gamma^\mu\gamma^5 \tau^3 Q$ to those of $\overline{Q} \gamma^\mu\gamma^5 (\tau^1\pm i\tau^2) Q$ which appear in the CC interaction. As a result, the values of $\tilde{C}_i^A(Q^2)=C_i^A(Q^2)$ can be derived from studying the CC resonant interaction \cite{Lalakulich:2005cs}.

\acknowledgments
The authors thank K. Azizi and M. Goharipour for useful discussions. 
MD thanks U. Mosel for helping us with utilizing GiBUU.
This work has received funding from the  Iran National Science Foundation (INSF) under project No. 4031487.
It  has also partially  been supported by the European
Union$'$s Framework Programme for Research and Innovation Horizon 2020 under grant H2020-MSCA-ITN2019/860881-HIDDeN as well as under the Marie Sklodowska-Curie Staff Exchange grant agreement No 5101086085-ASYMMETRY. 
YF would like to thank the University of Barcelona and its staff where a part of this work is done for their hospitality. 
MD and SS thanked IPM for their hospitality during their visit.

\section*{Code Availability Statement}
\label{sec:code}
All details of GiBUU that we use are accessible at \href{https://gibuu.hepforge.org/}{https://gibuu.hepforge.org/}. In addition, our modifications on GiBUU to take into account NC NSI are available at \href{https://github.com/dehpour/NC-NSI-GiBUU}{https://github.com/dehpour/NC-NSI-GiBUU}.

\bibliography{biblio}

\end{document}